%% file: brett_galla_submit.tex
\documentclass[twocolumn,showpacs,secnumarabic,preprintnumbers,amsmath,amssymb,10pt,a4paper,superscriptaddress,aps,pre,showpacs]{revtex4-1}
\usepackage{geometry}

\usepackage{graphicx}
\usepackage{dcolumn}
\usepackage{bm}
\usepackage{mathtools}
\usepackage{color}

\input{preamble.tex}

\newcommand{\uu}[1]{\mathbb{#1}}
\begin{document}

\title{Generating functionals and Gaussian approximations for interruptible delay reactions}

\author {Tobias Brett}
\email{tsbrett@umich.edu }
\affiliation{Theoretical Physics, School of Physics and Astronomy, The University of Manchester, Manchester M13 9PL, United Kingdom}
\affiliation{Department of Ecology and Evolutionary Biology, University of Michigan, Ann Arbor, MI 48109, USA}
\affiliation{Center for the Study of Complex Systems,\\ University of Michigan, Ann Arbor, MI 48109, USA}

\author{Tobias Galla}
\email{tobias.galla@manchester.ac.uk}
\affiliation{Theoretical Physics, School of Physics and Astronomy, The University of Manchester, Manchester M13 9PL, United Kingdom}

\begin{abstract}
We develop a generating functional description of the dynamics of non-Markovian individual-based systems, in which delay reactions can be terminated before completion. This generalises previous work in which a path-integral approach was applied to dynamics in which delay reactions complete with certainty. We construct a more widely applicable theory, and from it we derive Gaussian approximations of the dynamics, valid in the limit of large, but finite population sizes. As an application of our theory we study predator-prey models with delay dynamics due to gestation or lag periods to reach the reproductive age. In particular we focus on the effects of delay on noise-induced cycles.
\end{abstract}

\maketitle

\section{Introduction}

In recent years there has been a growth in the understanding of the importance of intrinsic noise in complex systems composed of a finite number of interacting constituents. It has been recognised that there are visible, macroscopic effects due to the stochasticity inherent in the interactions in a variety of systems, including for example metabolic pathways \cite{stelling.2004}, gene regulatory systems \cite{oudenaarden2001,oudenaarden2002,elowitz2002,raj.2008}, predator-prey dynamics \cite{bjornstad.2001,mckane} or models of disease spread \cite{rohani.2002,alonso.2007}. This inherent stochasticity is referred to as `intrinsic noise' or `demographic stochasticity' \cite{kaen}. Intrinsic noise can give rise to phenomena such as cyclic dynamics \cite{mckane, cremer.2008}, patterns and waves \cite{reichenbach.2007,reichenbach.2007b,butler,biancalani.2010,biancalani.2011}, and extinction events \cite{traulsen.2005,traulsen.2009,assaf.2010,ovaskainen.2010,mobilia.2010}. These phenomena are not captured by more traditional deterministic modelling approaches, instead they are purely noise-induced.

Deterministic models are built on ordinary or partial differential equations. These equations are formally only valid in the limit of an infinite system, that is the number of interacting constituents is so large that stochastic effects play no role \cite{vankampen}. To take into acount the intrinsic stochasticity of the interactions a full probabilistic description is required. Widely used modelling approaches are drawn from the theory of stochastic processes, most notably the master equation \cite{nordsieck, vankampen}, describing the time evolution of the probability distribution over the space of states. Formulating the master equation approach relies on the Markov property of the underlying dynamics: transition rates from one state to another must only depend on the target state and the present state of the system, but not on the path the dynamics has taken to arrive at the current state. This implies that the system has no memory of previous interactions, all effects of an interaction must be realised instantaneously. While this is a reasonable assumption for many processes in physics, this is typically not the case in biological models. For example birth in a predator-prey system occurs after a period of gestation, transcriptional and translational delays are relevant in gene regulatory systems \cite{monk.2003}, and recovery occurs a certain period of time after infection in models of disease spread \cite{anderson.1991,lloyd.2001a}. These processes can lead to situations in which effects of an initial interaction (e.g. impregnation, initiation of transcription, infection) materialise with a significant delay. Such dynamics are then no longer Markovian as the change of the state of the system at any one time $t$ may depend on what processes were set in motion in the past and which complete at $t$. Of course the modelling as a delay system can often be avoided through the construction of higher-dimensional models with intermediate states (e.g. staged models in epidemics \cite{lloyd.2001a}), but mathematically it is often convenient to use delay models, as these are relatively easy to set up for arbitrary delay distributions.

Traditionally delay dynamics have been investigated based on deterministic approaches \cite{yeung.1999,Michiels,milton.2009,monk.2003,Alboszta.2004,magpantay.2010}. These are subject to the limitations outlined above in that they do not capture the effects of stochasticity. It is only recently that analytical (and also numerical) techniques for individual-based models subject to both intrinsic noise and delay have been developed \cite{barrio,bratsun,anderson,cai,lafuerza.2011a,lafuerza.2011b,lafuerza.2012, galla,brett_galla,brett_galla2014,boguna.2014}. In previous work we have derived Gaussian approximations of such dynamics, and we have shown that these can capture the effects of delay reactions \cite{brett_galla,brett_galla2014}. This analysis is conveniently carried out in terms of generating functionals \cite{altland,zinn_justin}, and the equivalent of the linear-noise approximation (LNA) for delay systems can be derived. An alternative, more informal method to derive the same Gaussian approximations (by-passing the exact description) was highlighted in \cite{brett_galla2014}.

In these existing approaches delay reactions fire at an initial time with rates determined by the state of the system at that time. The reaction may then have an immediate effect on the composition of the system, and subsequently a second effect materialises at a later time (after the delay). The delay is either fixed, or it can be drawn from an underlying distribution each time a delay reaction is initiated. Existing work is often restricted to what we will refer to as `definitive' completion of the delay reaction. Once a delay reaction is initiated (it `fires') the delayed effect will occur, no matter what the trajectory of the system is between the time the reaction fires and the designated completion time. This is an obvious limitation of the modelling approach. Descriptions in which delay reactions may fail to complete depending on events that occur between initiation and designated completion provide a much more realistic description of many real-world processes. The gestation period in a predator-prey model presents perhaps the most intuitive example of delay reactions which can be interrupted.  A pregnancy is initiated at a given time and the birth event will occur at a designated later time. However, birth is not certain, the mother might die during pregnancy. Similarly, in a model of disease spread, an individual may get infected with the disease and would then be scheduled to recover at a later time. However, the individual may die in the interim, or be removed from the system in some other way, so that the completion event (recovery) may not occur. The rate with which such removal occurs may well depend on the state of the system at the time of removal.

The purpose of the present paper is to extend existing approaches to the modelling of stochastic dynamics with delay to the case in which delay reactions may not complete. We will refer to such reactions as `interruptible delay reactions'. In this setting delay reactions can be of several types: (i) Delay reactions which cannot be interrupted. This is the case considered in \cite{brett_galla}; (ii) Delay reactions which can be interrupted, but the probability of interruption does not depend on the state of the system. An example is death of the mother for internal reasons. Another example was also considered in \cite{brett_galla} in the context of the susceptible-infective-recovered (SIR) model with birth and death, with constant death rate; (iii)  Delay reactions which can be interrupted and the probability of interruption depends on the state of the system at the time of interruption. Continuing the example of a pregnant prey individual, the rate of predation depends on the number of predators present in the system throughout the pregnancy period. This case is \emph{not} covered by \cite{brett_galla}.

In this paper we develop a systematic Gaussian approximation to models with interruptible delay reactions. Specifically we extend the generating functional method used in \cite{brett_galla} to include interruption effects so that all three cases above are covered. As an application we consider the SIR model of disease spread and two variants of a predator-prey model, one with a delay period due to pregnancy and one with a delay period due to the maturation of juveniles.

\section{Model definitions: Interruptible delay reactions}
\label{sec:delay_uncert_theory}
\subsection{Model definition}
We consider a finite population of interacting constituents, each of which is of one of $M$ different types, labelled $\alpha = 1,\ldots,M$. We will refer to the constituents as `individuals' in the following. The state of the system at any one time is then described by an $M$-dimensional vector, $\bm{n}(t)=[n_1(t),\dots,n_M(t)]$, where the non-negative integer $n_\alpha(t)$ indicates the number of individuals of type $\alpha$ at time $t$. We assume a continuous-time evolution. The system is well-mixed and two individuals of the same type are indistinguishable. The individuals interact through a set of $R$ reactions, we will label these $i = 1,\ldots,R$. Each possible reaction $i$ is associated with an initiation rate $T_i(\bn)$, indicating the rate with which reactions of type $i$ fire if the system is in state $\bn$. When such a reaction fires an instantaneous change of the state of the system $\bm{n}(t)$ occurs at the time of firing, described by the vector $\bv_i$.  That is to say the state of the system changes from $\bn$ to $\bn+\bv_i$. Subsequently these reactions may also have a delayed effect. This is implemented as follows: if a delay reaction fires at time $t$ an instantaneous change of the state of the system occurs as described above. In addition to this, a delay time $\tau$ is drawn from an underlying delay distribution, $K_i(\tau)$. This distribution is specific to the reaction triggered, as indicated by the subscript $i$. After the delay time has elapsed a second change in the state of the population may occur. This happens at time $t+\tau$, and we denote the delayed effect of the reaction by $\bw_i$. The key element of the dynamics that we add in the present work is the possibility that a delay reaction may not complete. We assume that a delay reaction of type $i$, triggered at time $t$ and due to complete at $t+\tau$ can be interrupted at any time between $t$ and $t+\tau$. The rate with which this occurs is $f_i[\bn(t')]$, where $t<t'<t+\tau$. The termination rate may thus depend on the state of the system $\bn(t')$. If this happens the delayed effect at time $t+\tau$ does not occur, instead we assume that the state of the system changes by $\bu_i$ at time $t'$. For the time being we will only consider cases in which there is only one way in which each delay reaction can be interrupted. A generalisation to models in which delay reactions can be interrupted in multiple different ways is relatively straightforward, though slightly more cumbersome, see Appendix~\ref{sec:multiple_interruption} for details.

 As is commonly done in the modelling of interacting particle systems, we will assume that all reaction rates $T_i(\bn)$ scale with a parameter $\Omega$ -- that is to say $T_i(\bn)={\cal O}(\Omega)$. This parameter can be seen as setting the scale of the size of the system (the scale of the total number of individuals in the system, or equivalently the volume of the system), and the scaling of the rates reflects the fact that the total (average) number of reactions in the system per unit time is proportional to its size. For later purposes it is convenient to introduce the quantities $x_\alpha(t)=n_\alpha(t)/\Omega$. We will refer to these as the `concentrations' of particles of type $\alpha$.  We also introduce the intensive rates $r_i[\bm{x}(t)]=T_i[\bx(t)]/\Omega$.
 
\subsection{Discrete-time dynamics}
As in \cite{brett_galla} we will proceed by discretising time into steps of duration $\Delta$. The continuum limit will be restored at the end. We will write $\bx_{\alpha,t}$ for the concentration of individuals of type $\alpha$ at time step $t$. In the discretised model, we will assume that all reaction rates remain constant between $t$ and $t+\Delta$. If the concentration vector is $\bx_t$ at time $t$ then the number of newly triggered reactions of type $i$ during this time step is a Poissonian random variable $k_{i,t}$ with parameter $T_i(\bn)\Delta$. In the absence of delay reactions the dynamics of the discrete-time model would hence read
\begin{align}
x_{\alpha,t+\Delta} - x_{\alpha,t}= \frac{1}{\Omega}\sum_i  v_{i,\alpha} k_{i,t}.
\label{eq:change_in_x_no_delay}
\end{align}
In models with delay reactions we have to extend this expression to take into account (i) delay reactions completing at the designated time and (ii) delay reactions which terminate before the designated completion time. We then have
\begin{align}
 x_{t+\Delta,\alpha}-x_{t,\alpha} = \frac{1}{\Omega}\sum_i \sum_{\tau >0}\bigg[& v_{i,\alpha} k_{i,t}^{\tau} + w_{i,\alpha}m_{i,t-\tau}^{\tau} \nonumber \\
 &+ \sum_{0<s< \tau} u_{i,\alpha}\ell_{i,t-s}^{\tau,s} \bigg],
\label{eq:discretetime_eom}
\end{align}
In this expression $k_{i,t}^{\tau}$ is the number of reactions of type $i$ that fire at time $t$ and which have a delay period $\tau$. The quantity $m_{i,t-\tau}^{\tau}$ is the number of reactions of type $i$ that fired at time $t-\tau$ and successfully complete at time $t$. Finally, the quantity $\ell_{i,t-s}^{\tau,s}$ is the number of reaction of type $i$ that fired at time $t-s$ with a designated delay period $\tau$ (i.e. scheduled for completion at $t-s+\tau$), but which are interrupted at time $t$ ($0<s<\tau$). We note that non-delay reactions are included in this descriptions, they would simply have $\bw_i=\bu_i=0$.

 Eq.~\eqref{eq:discretetime_eom} defines the conditional probability 
\BE
 &&P(\bx_{t+\Delta}|\{\bx_t,\bm{k},\bm{\ell},\bm{m}\}_{t'\le t}) \nonumber \\
 &=& \prod_\alpha \delta\bigg(x_{t+\Delta,\alpha}-x_{t,\alpha} -\frac{1}{\Omega}\sum_i \sum_{\tau >0}\bigg[ v_{i,\alpha} k_{i,t}^{\tau} \nonumber \\
 &&+ w_{i,\alpha}m_{i,t-\tau}^{\tau} + \sum_{0<s< \tau} u_{i,\alpha}\ell_{i,t-s}^{\tau,s} \bigg]\bigg),
\label{eq:delta_eom}
\EE
where the notation $ \{\bx,\bm{k},\bm{\ell},\bm{m}\}_{t'\le t}$ indicates variables with indices $t' \le t$.

We have already described the Poissonian nature of the variables $k_{i,t}=\sum_\tau k_{i,t}^\tau$, but it remains to define in more detail how the $k_{i,t}^\tau$, $m_{i,t-\tau}^\tau$ and $\ell_{i,t-s}^{\tau,s}$ are chosen. This is explained in the following section.

\subsection{Statistics of reaction numbers}
We first discuss the statistics of the variables $k_{i,t}^\tau$, indicating the number of reactions of type $i$ triggered in the discrete time model at time step $t$ and with completion due at time $t+\tau$. As discussed in \cite{brett_galla} each $k_{i,t}^\tau$ is a Poissonian random variable with parameter $ \Delta^2K_i(\tau)\Omega r_i(\bx_t)$. This is not affected by possible interruptions of the delay reactions. At this point it is useful to recall the definition $r_i(\bm{x}_t)=T_i(\bx_t)/\Omega$. Broadly speaking $\Omega r_i(\bx_t)$ reflects the rate with which reactions of type $i$ are initiated (with any delay). Once a reaction is initiated a delay period is drawn independently from a distribution $K_i(\tau)$. In the discrete-time model this is reflected in the factor $\Delta K_i(\tau)$. 

Ultimately we will be interested in taking the continuous-time limit for the dynamics. Anticipating this, we focus on the case of small $\Delta$, which simplifies the problem significantly.  The probability distribution of $k_{i,t}^\tau$ is of the form
\begin{align}
P(k_{i,t}^\tau = 1|\bx_t) =&~ \Delta^2K_i(\tau)\Omega r_i(\bx_t)+\mathcal{O}(\Delta^4) ,\nonumber \\
P(k_{i,t}^\tau = 0|\bx_t) =&~ 1- \Delta^2K_i(\tau)\Omega r_i(\bx_t)+\mathcal{O}(\Delta^4),\nonumber \\
P(k_{i,t}^\tau > 1|\bx_t) =&~ \mathcal{O}(\Delta^4) .
\label{eq:prob_k}
\end{align}
Eq. (\ref{eq:prob_k}) indicates that the probability to observe two or more initiation events of reactions of type $i$ and with delay $\tau$ in any one time interval $\Delta$ is at least of order $\Delta^4$. In the limit of small $\Delta$ we can therefore restrict ourselves to $k_{i,t}^\tau \in \{0,1\}$. 

If $k_{i,t}^\tau = 0$ then it is clear that $\ell_{i,t}^{\tau,s} = 0$ for all $s$ and also $m_{i,t}^\tau=0$, i.e. we have
\begin{align}
P(\ell_{i,t}^{\tau,s} = 0|k_{i,t}^\tau=0) &= 1 \; \forall\; 0<s<\tau ,\nonumber \\
P(m_{i,t}^\tau = 0 | k_{i,t}^\tau=0) &= 1.
\label{eq:prob_k0}
\end{align}
If $k_{i,t}^\tau = 1$ then only one out of the $\{\ell_{i,t}^{\tau,s}\}_{0<s<\tau} $ and $m_{i,t}^\tau$ can be non-zero. If $k_{i,t}^\tau = 1$ and  $\ell_{i,t}^{\tau,\sigma} = 1$ then it follows that $\ell_{i,t}^{\tau,s} = 0$ for all $s\ne\sigma$, and $m_{i,t}^\tau=0$. I.e., we have
\begin{align}
P(\ell_{i,t}^{\tau,s} = 0|\ell_{i,t}^{\tau,\sigma}=1, k_{i,t}^\tau=1) &= 1 \; \forall\; \sigma<s<\tau, \nonumber \\
P(m_{i,t}^\tau = 0 |\ell_{i,t}^{\tau,\sigma}=1, k_{i,t}^\tau=1) &= 1.
\label{eq:prob_if_k1_l1}
\end{align}
As explained above, interruption happens with probability $\Delta \times f_i(\bx_{t+s})$ in the next time interval $\Delta$  if the system is in state $\bx_{t+s}$, and so
\BE
&&P(\ell_{i,t}^{\tau,s} = 1|\bm{x}_{t+s},\{\ell_{i,t}^{\tau,\sigma}=0\}_{0<\sigma<s}, k_{i,t}^\tau=1) \nonumber \\
&=& \Delta f_i(\bx_{t+s}),\nonumber\\ \nonumber \\
&&P(\ell_{i,t}^{\tau,s} = 0|\bm{x}_{t+s},\{\ell_{i,t}^{\tau,\sigma}=0\}_{0<\sigma<s}, k_{i,t}^\tau=1)\nonumber \\
&=& 1-\Delta f_i(\bx_{t+s}).
\label{eq:prob_l0_k1}
\EE
Finally, if the reaction has not been interrupted by time $t+\tau$ then the reaction always completes,
\begin{align}
 P(m_{i,t}^{\tau} = 1|\{\ell_{i,t}^{\tau,\sigma}=0\}_{0<\sigma<\tau}, k_{i,t}^\tau=1) &=  1.
\label{eq:prob_m_1}
\end{align}

\section{Generating functional}
\label{sec:uncertain_gf}

In discrete time the generating function is
\begin{equation}
Z(\bm{\psi}) = \avg{e^{-\Delta \sum_{t} \bm{\psi}_{t}\cdot \bm{x}_{t}}}_{\text{paths}},
\end{equation}
where the average is performed over all possible paths. We have introduced a source term ${\bm \psi}$, derivatives with respect to it generate correlation functions \cite{altland,zinn_justin} The main task in the calculation that follows is to compute the path-average in the above expression. This entails integrating/summing over all random variables, 
\BE
Z(\bm{\psi}) &= &\int \prod_{t}d \bm{x}_t \sum_{\bm{k},\bm{\ell},\bm{m}} \big[\mathcal{P}(\bm{x},\bm{k},\bm{\ell},\bm{m}) e^{-\Delta\sum_{t}\bm{\psi}_{t}\cdot \bm{x}_{t}}\big],\nonumber \\
\label{eq:gf_deff}
\EE
where $ \mathcal{P}(\bm{x},\bm{k},\bm{\ell},\bm{m})$ is the joint probability distribution of all $x_{t,\alpha}$, $k_{i,t}^\tau$, $\ell_{i,t}^{\tau,s}$, and $m_{i,t}^\tau$ -- i.e. the probability of a path. Our objective is to find an expression for the generating functional after performing the sums over all $\bm{k}$, $\bm{\ell}$, and $\bm{m}$ and after subsequently taking the continuous-time limit $\Delta\to 0$. The algebra involved is somewhat lengthy, and so we do not give full details here, they are relegated to Appendix~\ref{sec:gf_appendix}. The calculation consists of carrying out the following main steps:  (i) The path probability $\mathcal{P}(\bm{x},\bm{k},\bm{\ell},\bm{m})$ is expressed as a product of conditional probabilities using Eqs.~\eqref{eq:delta_eom}--\eqref{eq:prob_m_1}; (ii) The delta-functions in Eq. (\ref{eq:delta_eom}) are converted into their Fourier representations, in the process introducing the conjugate variables $\bp_t$;
(iii) Using Eqs.~\eqref{eq:prob_k}--\eqref{eq:prob_m_1} the combinations of variables with non-zero probability are identified, along with their statistical weight. Knowing the weight of each combination enables us to average over the  $\bm{k}$, $\bm{\ell}$, and $\bm{m}$; (iv) Finally, the continuous-time limit is restored. The resulting generating functional is

\begin{equation}
Z[\bm{\psi}] =  \int D\bx D\bp~ e^{-S[\bx,\bp] - \int dt \bm{\psi}(t) \cdot \bm{x}(t)},
\label{eq:interrupt_exact_gf}
\end{equation}
with the action
\BE
S[\bx,\bp]\! &=& \!- \!\!\int \!\! dt \left[ \bm{p}(t)\! \cdot  \!\dot{\bm{x}}(t) -\!\sum_i \!\left\{R_i^{(1)}(t)+R_i^{(2)}(t)\right\}\right]\!. \nonumber \\
\label{eq:interrupt_hamiltonian}
\EE
There are two contributions to the action from each type of reaction: (i) a contribution from the reactions which are not interrupted,
\BE
  -R_i^{(1)}(t)=\int_0^\infty d\tau &\bigg(e^{ - \frac{1}{\Omega}\left[ \bm{v}_{i} \cdot \,\bm{p}(t-\tau) + \bm{w}_{i} \cdot\, \bm{p}(t) \right]} - 1 \bigg) \nonumber \\ &\hspace{-7em}\times e^{-\int_0^\tau d\sigma f_i[\bx(t-\sigma)]}K_i(\tau)\Omega r_i[\bx(t-\tau)],
\label{eq:no_interrupt_contib}
\EE
and (ii) a contribution from the reactions which are interrupted,
\BE
-R_i^{(2)}(t) &=&\int_0^\infty ds\bigg( e^{ - \frac{1}{\Omega}\left[ \bm{v}_{i} \cdot \, \bm{p}(t-s)  + \bm{u}_{i} \cdot \, \bm{p}(t) \right]}-1\bigg) \nonumber \\
&&\hspace{-5em}\times f_i[\bx(t)]e^{-\int_0^s d\sigma f_i[\bx(t-\sigma)]} \int_s^\infty d\tau K_i(\tau)\Omega r_i[\bx(t-s)].\nonumber \\
\label{eq:interrupt_contrib}
\EE
These contributions are both functionals of the path of $\bm{x}$ between the time at which reaction first fires and the time it either completes successfully or is interrupted. 

We notice the factor $e^{-\int_0^\tau d\sigma f_i[\bx(t-\sigma)]}$ in Eq.~\eqref{eq:no_interrupt_contib}. This exponential is the  probability that a delay reaction of type $i$ triggered at $t-\tau$ and with completion time $t$, reaches completion, given a path $\bx$ between $t-\tau$ and $t$. Similarly, the quantity $f_i[\bx(t)]e^{-\int_0^s d\sigma f_i[\bx(t-\sigma)]}dt$ [cf. Eq.~\eqref{eq:interrupt_contrib}] indicates the probability that a delay reaction of type $i$ triggered at time $t-s$ is interrupted in the time interval $[t,+dt)$, given a path $\bx$ between $t-s$ and $t$. 

If all delay reactions of type $i$ complete with certainty (i.e. in absence of interruptions) one has $f_i(\bx) = 0$. This implies $R_i^{(2)}=0$ (see Eq.~\eqref{eq:interrupt_contrib}). The quantity $R_i^{(1)}(t)$ in Eq.~\eqref{eq:no_interrupt_contib} reduces to the corresponding object in \cite{brett_galla}.

\section{Application to the SIR model with birth and death} \label{sec:interrupt_sir}

The SIR model with birth and death can be studied using the approach developed in Sec.~\ref{sec:delay_uncert_theory}.  This model describes the dynamics of an infectious disease in a population of individuals. Each individual can be in one of three states: susceptible, infectious or recovered. We consider an infection dynamics with delayed recovery. Upon contact with an infectious individual a susceptible member of the population may become infectious (with rate $\beta$), and then they recover at a time $\tau$ after infection. The delay time $\tau$ is drawn from a recovery-time distribution $K(\tau)$. We write this process as follows
\be
S+I\overset{\beta}{\longrightarrow} 2I; I\overset{K(\tau)}{\Longrightarrow} R.
\label{eq:sir_reactions1}
\ee
The first arrow indicates the effect the reaction has at the time it is initiated. The double arrow indicates the delayed effect $\tau$ units of time after the reaction is triggered.  In addition to the recovery process, any individual may die, this occurs at constant rate $\mu$, independent of the infection status of the individual. In order to keep the population size, $N$, constant any dying individual is immediately replaced by an individual of type $S$. This leads to the following additional reactions
\BE
I \overset{\mu}{\longrightarrow} S, \nonumber \\
R \overset{\mu}{\longrightarrow} S. \label{eq:sir_reactions2}
\EE
Crucially, an individual may die (and be replaced by an $S$) after becoming infected, but before the scheduled recovery time. This represents a delay reaction with possible interruption in the formalism described in the previous sections. In this introductory example, however, the rate of interruption does not depend on the state of the system and so this model can be studied using the simpler method presented \cite{brett_galla}. There we mapped the above reaction scheme onto the model
\begin{align}
R &\stackrel{\mu}{\longrightarrow} S, \nonumber \\
S+I&\stackrel{\chi\beta}{\longrightarrow} 2I; I\stackrel{\bar K(\tau)}{\Longrightarrow} R, \nonumber \\
S+I&\stackrel{(1-\chi)\beta}{\longrightarrow} 2I; I\stackrel{\bar Q(s)}{\Longrightarrow} S,
\label{eq:sir_splitreaction_method}
\end{align}
with
\begin{align}
\chi &=\int_0^\infty d\tau K(\tau)\int_\tau^\infty ds~ \mu e^{-\mu s} ,  \nonumber \\
\bar K(\tau) &= \chi^{-1}K(\tau)\int_{\tau}^\infty ds~ \mu e^{-\mu s}, \nonumber \\
\bar Q(s) &= (1-\chi)^{-1}\mu e^{-\mu s} \int_{s}^\infty d\tau~ K(\tau). \label{eq:Q_sir_interrupt}
\end{align}
The first line in Eq. (\ref{eq:sir_splitreaction_method}) describes the standard death and replacement reaction $R \stackrel{\mu}{\longrightarrow} S$. The second and third lines represent two reaction sequences which may be triggered when an infection event occurs. Each sequence starts with an infection event (occurring with rate $\beta n_I n_S/N$). With probability $\chi$ the newly infected individual is scheduled for recovery at a later time; the time-to-recovery, $\tau$, is drawn from the distribution $\bar K(\tau)$. With complementary probability $1-\chi$ the newly infected individual is scheduled for death (and replacement by an individual of type $S$). The time-to-replacement, $s$, is drawn from the distribution $\bar Q(s)$. This mapping is possible because the interruption rate, $\mu$, is independent of the state of the system. Both distributions, $\bar{K}$ and $\bar{Q}$, are normalised.

The reaction scheme in Eq.~(\ref{eq:sir_splitreaction_method}) and the distributions in Eq. (\ref{eq:Q_sir_interrupt}) were formulated in \cite{brett_galla}, and from this scheme the generating functional of the process was derived. Now that we have put the more general formalism of the previous sections in place we can recognise these existing results as a special case.  Inserting the model specifications into the general formalism of the previous sections the action in Eq. (\ref{eq:interrupt_hamiltonian}) is indeed found as
\BE
S[\bx,\bp] &=& \int dt \Bigg\{ p_S(t)\dot{x}_S(t) +  p_I(t)\dot{x}_I(t) \nonumber \\ 
&&\hspace{-3em}+\bigg(e^{ - \frac{1}{N}p_S(t)} - 1 \bigg)N(1-x_S(t)-x_I(t)) \nonumber \\
&&\hspace{-3em}+\int_{-\infty}^t dt' \bigg(e^{ - \frac{1}{N}\left[ p_I(t')- p_S(t') - p_I(t) \right]} - 1 \bigg) \nonumber \\ 
&&\hspace{-3em}\times \bar K(t-t') N \chi \beta x_S(t')x_I(t') \nonumber \\
&&\hspace{-3em}+\int_{-\infty}^t dt'\bigg( e^{ - \frac{1}{N}\left[  p_I(t')- p_S(t') - p_I(t) + p_S(t) \right]}-1\bigg) \nonumber \\
&&\hspace{-3em}\times \bar Q(t-t')N (1-\chi)\beta x_S(t')x_I(t') \Bigg\},
\label{eq:SIR_hamiltonian}
\EE

where we have used the definitions of $\chi$, $\bar K(\tau)$, and $\bar Q(s)$ as given in Eqs.~\eqref{eq:Q_sir_interrupt}. The total size of the population, $N =n_S + n_I +  n_R$ is constant in time, so there are only two independent degrees of freedom. In formulating Eq. (\ref{eq:SIR_hamiltonian}) we have therefore eliminated species $R$ via $n_R = N- n_S - n_I$ (equivalently $x_R=1-x_S-x_I$). It is important to stress that we have used $\Omega=N$ for simplicity, given the constant size of the population. The result in Eq. (\ref{eq:SIR_hamiltonian}), derived from the general formalism of the previous sections, is the same action as the one found in \cite{brett_galla} using the mapping of Eq.~\eqref{eq:sir_splitreaction_method}.

\section{Approximations to the generating functional}
\subsection{Deterministic approximation}

The exact generating functional can be approximated by Taylor expanding the action in powers of the inverse system size. To leading order in $\Omega^{-1}$ the general action Eq.~\eqref{eq:interrupt_hamiltonian} reduces to
\be
S[\bx,\bp] = -\!\int\! dt \Bigg[ \bm{p}(t) \cdot \dot{\bm{x}}(t)-\sum_i\bigg\{A_i^{(1)}(t)+A_i^{(2)}(t) \bigg\}\Bigg],
\ee
where
\BE
A_i^{(1)}(t)&=&\int_0^\infty d\tau \bigg( \bm{v}_{i} \cdot \,\bm{p}(t-\tau) + \bm{w}_{i} \cdot\, \bm{p}(t) \bigg) \nonumber \\
&&\times e^{-\int_0^\tau d\sigma f_i[\bx(t-\sigma)]}K_i(\tau) r_i[\bx(t-\tau)],\nonumber \\
\EE
and
\BE
A_i^{(2)}(t)&=&\int_0^\infty ds\bigg( \bm{v}_{i} \cdot \, \bm{p}(t-s)  + \bm{u}_{i} \cdot \, \bm{p}(t)\bigg) \nonumber \\
&&\times f_i[\bx(t)]e^{-\int_0^s d\sigma f_i[\bx(t-\sigma)]} \nonumber \\
&&\times \int_s^\infty d\tau K_i(\tau) r_i[\bx(t-s)].
\EE
This action is linear in $\bp$ (by construction), and it describes the deterministic dynamics 
\begin{align}
\dot{x}_\alpha^\infty = \sum_i v_{i,\alpha}r_i[\bx^\infty(t)] +F_\alpha^{(1)}[\bx^\infty](t)+F_\alpha^{(2)}[\bx^\infty](t), \label{eq:det_approx}
\end{align}
with the drift terms
\begin{align}
F_\alpha^{(1)}[\bx^\infty](t) = \sum_i \!\bigg[& \int_0^\infty \!d\tau\; w_{i,\alpha}e^{-\int_0^\tau d\sigma f_i[\bx^\infty(t-\sigma)]} \nonumber \\ 
&~~\times K_i(\tau)r_i[\bx^\infty(t-\tau)]\bigg] ,
\end{align}
and
\begin{align}
 &F_\alpha^{(2)}[\bx^\infty](t) =\sum_i\bigg[ \int_{0}^\infty \!ds \; u_{i,\alpha}f[\bx^\infty(t)] \nonumber \\
&~~~~~~\times e^{-\int_0^s \!d\sigma f_i[\bx^\infty(t-\sigma)]} \int_s^\infty d\tau K_i(\tau)r_i[\bx^\infty(t-s)]\bigg],
\end{align}
featuring due to the delay. The deterministic approximation is valid for large (formally infinite) system sizes, and neglects all stochastic effects. We have denoted the dynamical variables in the deterministic limit by $\bx^\infty$. 

\subsection{Linear-noise approximation}

The linear-noise approximation is used to study fluctuations about the solution to the deterministic equations of motion, Eq.~\eqref{eq:det_approx}. We write $\bx= \bx^\infty + \Omega^{-1/2}\bm{\xi}$, and formulate the generating functional in terms of the variable $\bm{\xi}$. In doing this we re-scale the conjugate variables via $\bp \to \sqrt{\Omega}\bm{q}$. The resulting generating functional is
\begin{align}
Z[\bm{\psi}] =&  \int D\boldxi D\bm{q} ~e^{-S[\bx^\infty + \Omega^{-1/2}\bm{\xi}, \sqrt{\Omega}\bm{q}]} \nonumber \\
&\times  e^{- \int\! dt \,\bm{\psi}(t) \cdot \left\{\bm{x}^\infty(t) + \Omega^{-1/2}\bm{\xi}(t)\right\}}.
\label{eq:gf_uncert_sub}
\end{align}
Differentiating Eq.~\eqref{eq:gf_uncert_sub} with respect to $\bm{\psi}(t)$ still gives the moments and correlation functions of $\bm{x}(t)$. To find a generating functional for $\boldxi$ we divide Eq.~\eqref{eq:gf_uncert_sub} by the factor $e^{- \int dt~ \bm{\psi}(t)\cdot \bm{x}^\infty(t)}$, which is independent of $\boldxi$ and $\bm{q}$, and rescale the source term, $\bm{\psi} \to \sqrt{\Omega} \bm{\phi}$. The generating functional for $\boldxi$ is
\begin{equation}
Z_\xi[\bm{\phi}] =  \int D\boldxi D\bm{q} ~e^{-S[\bx^\infty + \Omega^{-1/2}\bm{\xi}, \sqrt{\Omega}\bm{q}] - \int dt \bm{\phi}(t)\cdot \bm{\xi}(t)}.
\label{eq:gf_xi_uncert}
\end{equation}
The moments of $\boldxi$ can now be found by differentiating $Z_\xi[\bm{\phi}]$ with respect to $\bm{\phi}$. 

The terms in the action are again expanded in powers of $\Omega^{-1/2}$, and the expansion is curtailed after sub-leading order (i.e. keeping terms $\mathcal{O}(\Omega^{0})$ and above). In order to illustrate the procedure we will consider the case where the deterministic dynamics are at a stable fixed point, $\bx^\infty(t) = \bx^*$. This is not necessary for the linear-noise approximation, however it will allow us to keep the amount of algebra under control. The action is $ S[\bx^* + \Omega^{-1/2}\bm{\xi}, \sqrt{\Omega}\bm{q}] = S_{\text{LNA}}[\bm{\xi},\bm{q}] + \mathcal{O}(\Omega^{-1/2})$ with
\BE
 S_{\text{LNA}}[\bm{\xi}, \bm{q}] &=& -\int dt\sum_\alpha q_\alpha(t)\dot{\xi}_\alpha(t)  \nonumber \\
 &&\hspace{-6em}+ \int dt \int dt'  \sum_{\alpha,\beta}q_\alpha(t)A_{\alpha,\beta}[\bx^*](t,t') \xi_\beta(t') \nonumber \\
&&\hspace{-6em}+ \frac{1}{2}\int dt \int dt' \sum_{\alpha,\beta} q_\alpha(t) B_{\alpha,\beta}[\bx^*](t,t')q_\beta(t') , 
\label{eq:action_lna_uncert}
\EE
with $A_{\alpha,\beta}[\bx^*](t,t') =\left. \frac{\delta \dot{x}_\alpha^\infty(t)}{\delta x_\beta^\infty(t')}\right|_{\bx^\infty = \bx^*}$. In this expression $\dot x_\alpha^\infty$ stands for the right-hand side of Eq. (\ref{eq:det_approx}).

The action is now quadratic in $\bm{q}$ and $\bm{\xi}$, and it describes a process with additive Gaussian coloured noise \cite{zinn_justin, aron}. This process is of the form
\begin{align}
\dot{\xi}_\alpha(t) =  \sum_\beta \int d t' A_{\alpha,\beta}[\bx^*](t,t')\xi_\beta(t') + \eta_\alpha(t),
\label{eq:lna_langevin_uncert}
\end{align}
with noise correlator $\avg{\eta_\alpha(t)\eta_\beta(t')} = B_{\alpha,\beta}[\bm{x}^*](t,t')$. The explicit expressions for $A_{\alpha,\beta}[\bx^*](t,t')$ and $B_{\alpha,\beta}[\bx^*](t,t')$ are somewhat lengthy, and before presenting them it is helpful to make a few simplifying definitions. 
\subsection{Further simplications}
To simplify the above expressions further we define the distributions
\begin{align}
\bar{K}_i(\tau) &= \frac{1}{\chi_i(\bm{x}^*)}e^{-f_i(\bm{x}^*)\tau}K_i(\tau) , \nonumber \\
\bar{Q}_i(s) &= \frac{1}{1-\chi_i(\bm{x}^*)}f_i(\bm{x}^*)e^{-f_i(\bm{x}^*)s}\int_{s}^\infty d\tau K_i(\tau),
\label{eq:lna_distributions}
\end{align}
for $\tau, s > 0$. We set $\bar{K}_i(\tau) =0$ for $\tau\leq 0$, and $\bar{Q}_i(s) = 0$ for $s\leq 0$, and we introduce $\chi_i(\bm{x}^*) = \int_0^\infty d\tau e^{-f_i(\bm{x}^*)\tau}K_i(\tau)$. Within the linear-noise approximation and assuming that the deterministic dynamics are at the fixed point ${\bm x^*}$ this is the probability that a delay reaction of type $i$, once triggered, completes. The function $\bar K_i(\tau)$ is the conditional time to completion. The quantity $1-\chi_i(\bm{x}^*)$, on the other hand is the probability that the delay reaction is interrupted before completion, and $\bar Q_i(\tau)$ describes the conditional distribution of times to interruption. Both distributions $\bar{K}_i(\tau)$ and $\bar{Q}_i(\tau)$ are normalised. One notices that the expressions in Eq. (\ref{eq:Q_sir_interrupt}) are special cases of Eq. (\ref{eq:lna_distributions}), with $f_i(x^*)\equiv \mu$. We stress though that our analysis below will explicitly account for (linear) fluctuations of the interruption rate as the state of the system varies in time.  These fluctuations of the interruption rate are obviously not present in the system with constant $f_i(\bx)=\mu$, considered in \cite{brett_galla}.
\\

Additionally it is helpful to make the definition 
\BE
L_{i,\alpha}(\tau) &=& \chi_i(\bm{x}^*)\bar{K}_i(\tau)w_{i,\alpha}\nonumber \\
&& +[1-\chi_i(\bm{x}^*)]\bar{Q}_i(\tau)u_{i,\alpha}.
\EE
This quantity can be interpreted as the weighted delay effects (due to either successful completion or interruption) on species $\alpha$ of reactions of type $i$ a time period $\tau$ after such a reaction was triggered. 

Using these definitions we can write $A_{\alpha,\beta}[\bx^*](t,t')$ concisely as 
\BE
A_{\alpha,\beta}[\bx^*](t,t') &=&  \sum_i\bigg\{ v_{i,\alpha}\frac{\partial r_i(\bx^*)}{\partial x_\beta^*}\delta(t-t')  \nonumber \\
&&\hspace{-3em}+  u_{i,\alpha}[1-\chi_i(\bm{x}^*)]\frac{r_i(\bm{x}*)}{f_i(\bx^*)}\frac{\partial f_i(\bx^*)}{\partial x_\beta^*}\delta(t-t') \nonumber \\
&&\hspace{-3em}-  \int_{t-t'}^\infty d\tau L_{i,\alpha}(\tau) \frac{\partial f_i(\bx^*)}{\partial x_\beta^*}r_i(\bm{x}^*) \Theta(t-t') \nonumber \\
&&\hspace{-3em}+   L_{i,\alpha}(t-t')\frac{\partial r_i(\bm{x}^*)}{\partial x_\beta^*}\bigg\}.
\label{eq:A_uncert}
\EE

Each of the four terms in the curly brackets has a clear interpretation. As the system fluctuates around the fixed point, the number of reactions triggered, interrupted and completed will fluctuate as well. The first term is the contribution to $\dot{\xi}_{\alpha}(t)$ of initial effects of reactions (at the time of triggering) due to fluctuations. The second term is the contribution of interruption effects due to fluctuations at the time of interruption. The third term is the contribution of delayed effects (either successful completion or interruption) due to fluctuations between the reaction firing and the delayed effects occurring. Finally, the fourth term is the contribution of delayed effects due to fluctuations at the time the reaction first fired.

If the rate of interruption, $f_i$, is independent of the state of the system for all $i$ then only the first and last terms in Eq.~\eqref{eq:A_uncert} are non-zero.


We now turn to the correlation $B_{\alpha,\beta}[\bx^*](t,t')$ of the Gaussian noise variables in the linear-noise approximation. Given that delay reactions can have effects on the composition of the population at multiple times, this noise is not white. Instead one finds
\BE
B_{\alpha,\beta}[\bm{x}^*](t,t') &=& \sum_{i} r_i(\bm{x}^*)\bigg\{ \Big[ v_{i,\alpha}v_{i,\beta}+w_{i,\alpha}w_{i,\beta}\chi_i(\bm{x}^*) \nonumber \\
&& \hspace{-2em}+ u_{i,\alpha}u_{i,\beta} [1-\chi_i(\bm{x}^*)] \Big]\delta(t-t')\nonumber \\
&& \hspace{-2em}+  v_{i,\alpha}L_{i,\beta}(t'-t)+   v_{i,\beta}L_{i,\alpha}(t-t') \bigg\}.
\label{eq:lna_noise_concise}
\EE

As $L_{i,\alpha}(t) = 0$ for $t\leq 0$ only one of the three terms in Eq.~\eqref{eq:lna_noise_concise} is non-zero for any pair of times $t$ and $t'$ and any given reaction $i$.

\subsection{Spectrum of fluctuations}
The linear-noise approximation can be used to calculate the Fourier spectrum of fluctuations about the deterministic fixed point \cite{mckane}. This is particularly useful to study noise-induced quasi cycles, as we will explain in further detail below. Fourier transforming Eq.~\eqref{eq:lna_langevin_uncert}, gives
\begin{align}
 i\omega&\widetilde \xi_\alpha(\omega) = \sum_{\beta,i}\bigg\{\frac{i}{\omega}\Big(w_{i,\alpha}\chi_i(\bm{x}^*) - \widetilde L_{i,\alpha}(\omega) \nonumber \\
 &+u_{i,\alpha}[1-\chi_i(\bm{x}^*)]\Big) \frac{\partial f_i(\bm{x}^*)}{\partial x_\beta^*}r_i(\bm{x}^*)\nonumber \\
&+ \left(v_{i,\alpha}+ \widetilde L_{i,\alpha}(\omega)\right)\frac{\partial r_i(\bm{x}^*)}{\partial x_\beta^*} \nonumber \\
&+ u_{i,\alpha}[1-\chi_i(\bm{x}^*)]\frac{r_i(\bm{x}^*)}{f_i(\bm{x}^*)} \frac{\partial f_i(\bm{x}^*)}{\partial x_\beta^*} \bigg\}\widetilde\xi_\beta(\omega) + \widetilde\eta_\alpha(\omega).
\label{eq:interrupt_lna_langevin_ft}
\end{align}

This equation is linear in $\widetilde{\bm{\xi}}$, and it can be written as $\widetilde {\bm{\eta}}(\omega) = \uu{M}(\omega)\widetilde{\bm{\xi}}(\omega)$, with a suitable matrix $\uu{M}(\omega)$. The Fourier transform of Eq.~\eqref{eq:lna_noise_concise} (with respect to $t-t'$) is
\begin{align}
\widetilde B_{\alpha,\beta}(\omega) = \sum_{i} r_i(\bm{x}^*)\bigg\{& v_{i,\alpha}v_{i,\beta}+w_{i,\alpha}w_{i,\beta}\chi_i(\bm{x}^*) \nonumber \\
&+ u_{i,\alpha}u_{i,\beta} [1-\chi_i(\bm{x}^*)] \nonumber \\
& +  v_{i,\alpha}\widetilde L_{i,\beta}^*(\omega)+   v_{i,\beta}\widetilde L_{i,\alpha}(\omega) \bigg\}.
\end{align}
The spectrum of fluctuations about the deterministic fixed point is then characterised by the matrix $\uu{S}(\omega) = \avg{\widetilde{\bm{\xi}}(\omega)\widetilde{\bm{\xi}}^\dagger(\omega)}$ and it can be found from (see \cite{gardiner})
\begin{equation}
\uu{S}(\omega) = \uu{M}(\omega)^{-1}\uu{B}(\omega)[\uu{M}^\dagger(\omega)]^{-1}.
\label{eq:spectrum_matrix}
\end{equation}

\section{Application to a predator-prey model}

\label{sec:pred_prey}
\subsection{Model definitions}
As an application of the above formalism we will now consider stochastic effects in a predator-prey model with different types of delay. Specifically we write $X$ to denote prey-individuals, and $Y$ for predators. We focus on the dynamics governed by the following reactions
\BE
X &\overset{b(1-x/k)}\longrightarrow &2X, \nonumber \\
Y+X &\overset{p}\longrightarrow& 2Y ,\nonumber \\
Y &\overset{d}\longrightarrow& \emptyset.
\EE
We write $n_X$ and $n_Y$ for the number of individuals of each type, and $x=n_X/\Omega, y=n_Y/\Omega$.  As before $\Omega$ is a parameter, setting the scale of the population size. The first reaction describes reproduction of prey with a logistic birth rate, dependent on the concentration $x$. The constant $k$ represents a carrying capacity, more precisely, the system can contain a most $k\Omega$ individuals of the prey type.  The logistic birth rate for prey distinguishes this model from other stochastic predator-prey models \cite{mckane}. Within our stylised approach we assume that predation events result in the birth of a predator (second reaction),  the third reaction finally describes a death process for predators.

This is obviously a minimalist model, but it is in line with previous stylised modelling approaches for birth-death processes \cite{kendall.1948}, and it serves as a helpful testbed. 

There are various ways in which delay processes can be introduced into this model. One is to include gestation periods, in which a prey enters a pregnant state before giving birth. After the gestation period the pregnant individual returns to the regular (non-pregnant) state and a new prey individual is created.  Another possibility is to assume that newly born prey individuals are initially in a  `juvenile' state and that they cannot immediately reproduce. After a maturation period they become full adult prey individuals and acquire the ability to reproduce. We will refer to the first modification as the `gestation model' and the second modification as the `juvenile model'. In both models there is an additional intermediary class of individuals, denoted $X'$ -  pregnant prey in the gestation model and juvenile prey in the juvenile model. The class $X$ corresponds to non-pregnant prey in the gestation model and adult prey in the juvenile model.  We write $n_{X'}$ for the number of individuals of type $X'$ and and $ x'=n_{X'}/\Omega$. Of course one could introduce a similar state in the context of predators, but in-line with the above stylised approach we keep the complexity of the model to a minimum.

For the gestation model the reactions are
\begin{align}
X &\overset{b(1-x_{\rm tot}/k)}\longrightarrow X' \,;\, X' \overset{K(\tau)}\Longrightarrow 2X ,\nonumber \\
Y+X &\overset{p}\longrightarrow 2Y ,\nonumber \\
Y &\overset{d}\longrightarrow \emptyset.\label{eq:gest}
\end{align}
where we have introduced $x_{\rm tot} = x + x'$, the total concentration of prey. In the first reaction, a non-pregnant individual becomes pregnant with rate $b(1-x_{\rm tot}/k)$. They then give birth after a delay drawn from $K(\tau)$ to a non-pregnant individual and return to the non-pregnant state. 

For the juvenile model we use
\begin{align}
X &\overset{b(1-x_{\rm tot}/k)}\longrightarrow  X+ X' \,;\, X' \overset{K(\tau)}\Longrightarrow X, \nonumber \\
Y+X &\overset{p}\longrightarrow 2Y ,\nonumber \\
Y &\overset{d}\longrightarrow \emptyset.\label{eq:juve}
\end{align}
In this model an adult prey gives birth to a juvenile individual with rate $b(1-x_{\rm tot}/k)$. After a delay drawn from $K(\tau)$ the juvenile becomes an adult individual. 

Additionally, in both models, the intermediary individuals, $X'$, can be predated upon, via
\begin{equation}\label{eq:interrupt}
Y+X'  \overset{p}\longrightarrow 2Y.
\end{equation}
It is this predation reaction that leads to the possibility of interrupting a delay reaction before its scheduled completion. If a pregnant individual is eliminated in a predation event, it will obviously not give birth at the designated time, and similarly if a juvenile individual is removed during predation it will not reach its reproductive age. Crucially, the rate with which such events happen depends on the state of the system (e.g. on the number of predators in the population at that time). The methods of \cite{brett_galla} are hence not applicable, and we will instead use the extended and more general formalism developed earlier in this paper.

Both models have the same reaction rates. The only difference between the two models lies in the changes of $n_X$ and $n_{X'}$ when a delay reaction triggers and completes, i.e. in the numerical values of the stoichiometric coefficients $v_{1,x}$ and $w_{1,x}$ (we label the delay reaction by $i=1$ in both models). It is therefore convenient to carry out the analysis for both models together, keeping $v_{1,x}$ and $w_{1,x}$ general. After performing the calculations we will substitute in for these parameters and compare the two models.

Simulations of the models can be performed using the Modified Next Reaction Method (MNRM) \cite{cai,anderson}. The interruption reaction shown in Eq.~(\ref{eq:interrupt}) fires like a conventional reaction in the MNRM algorithm with rate $f_1[\bx(t)]\times m_1(t)$ where $m_1(t)$ is the number of `active' delay reactions at time $t$, i.e. reactions of type $i=1$ which have fired but for which the delayed effects have not yet occurred. When the interruption reaction occurs an element of the list of queued delay reactions is selected at random with uniform weights, and removed from the list. The effects of the interruption reaction are then applied to the state of the system according to $u_{1,\alpha}$. Aside from this the MNRM is unchanged.\\

\subsection{Deterministic dynamics}

Both models describe three species, $X, X'$ and $Y$. The primary reactions are listed in Eq. (\ref{eq:gest}) and (\ref{eq:juve}) respectively. These are labelled $i=1, 2, 3$ from top to bottom. The reactions rates for both systems are
\begin{equation}
\br [\bx(t)] =  \left( \begin{array}{c}
x(t)h[x_{\rm tot}(t)] \\
p x(t)y(t) \\
dy(t)
\end{array} \right),
\end{equation}
with $h(x_{\rm tot}) = b(1-x_{\rm tot}/k)$. The first reaction, $i=1$, is the only delay reaction. This reaction can be interrupted due to predation on the intermediaries, with rate $f_1[\bx(t)] = p y(t)$ per intermediary (i.e. any instance of the delay reaction which is active at time $t$ is subject to interruption with rate $py(t)$). The delay distribution is $K(\tau)$. The stoichiometric coefficients, $v_{i,\alpha}$, describing changes to the system when reactions trigger can be summarised as follows,
\begin{equation}
\underline{\underline{v}} =  \left( \begin{array}{ccc}
v_{x} & 1 & 0 \\
-1 & 0 & 1 \\
0 & 0 & -1
\end{array} \right),
\end{equation}
where the rows each stand for one reaction ($i=1,2,3$) and the columns represent the three types of individuals ($X, X'$ and $Y$). The only non-zero stochiometric coefficients for interruption events [see Eq. (\ref{eq:interrupt})] are $u_{1,x_{\rm tot}} = -1$ and $u_{1,y} = 1$. The non-zero stochiometric coefficients for successful completion of delay reactions are $w_{1,x}= w_{x}$ and $w_{1,x'}= -1$. The differences between the two models are in the numerical values of $v_{1,x}=v_x$ and $w_{1,x}=w_x$. For the juvenile model we have $v_{x}= 0, w_{x} = 1$, for the gestation model we have $v_{x} = -1, w_{x}  = 2$.

The deterministic equations of motion are found as
\BE
 \dot{x}^\infty(t) &=&v_x x^\infty(t)h[x_{\rm tot}^\infty(t)]- px^\infty(t)y^\infty(t) \nonumber \\
&&\hspace{-2em}+w_x \int_0^\infty d\tau e^{-\int_0^\tau d\sigma py^\infty(t-\sigma)} \nonumber \\
&&\hspace{-2em}\times K(\tau) x^\infty(t-\tau)h[x_{\rm tot}^\infty(t-\tau)] , \label{eq:pp_deterministic_x} \\ \nonumber \\
 \dot{y}^\infty(t)& =& px^\infty(t)y^\infty(t) -dy^\infty(t) \nonumber \\
&&\hspace{-2em}+ \int_0^\infty d\tau ~p y^\infty(t)e^{-\int_0^\tau d\sigma py^\infty(t-\sigma)} \nonumber \\ 
&&\hspace{-2em}\times \int_\tau^\infty ds K(s)x^\infty(t-\tau)h[x_{\rm tot}^\infty(t-\tau)],
\label{eq:pp_deterministic_y}
\EE
with $x_{\rm tot}^\infty = x^\infty + x'^\infty$. In principle we can also write an equation for $\dot{x}'^\infty$, however we know from the reactions that the number of $X'$ at any one time $t$ is equal to the total number of $X'$ created up to that point (through birth) and which have not been removed through maturation or predation. We find
\begin{align}
 x'^\infty(t) =& \int_0^\infty d\tau ~\bigg\{ e^{-\int_0^\tau d\sigma py^\infty(t-\sigma)} \nonumber \\ 
 &\times \int_\tau^\infty ds ~K(s)x^\infty(t-\tau)h[x_{\rm tot}^\infty(t-\tau)]\bigg\}.
\label{eq:conc_x'}
\end{align}

\subsection{Fixed point analysis}

At the fixed point the left-hand sides of Eqs.~\eqref{eq:pp_deterministic_x} and \eqref{eq:pp_deterministic_y} are zero, the concentration of intermediary individuals can be found from Eq.~\eqref{eq:conc_x'}. We obtain
\begin{align}
0 =& [w_x \chi(y^*)  + v_x ]x^*h(x_{\rm tot}^*)- px^*y^* ,\label{eq:pp_eom_x_fp} \\
0=& ~ px_{\rm tot}^*y^* -dy^*, \label{eq:pp_eom_y_fp} \\
0 =&~ [1-\chi(y^*)]x^*h(x_{\rm tot}^*) - px'^*y^*, \label{eq:pp_eom_xp_fp}
\end{align}
where the stars indicate the fixed-point values of the relevant variables. We have also introduced the quantity $\chi(y^*) = \int_0^\infty d\tau~ e^{-py^*\tau}K(\tau)$, representing the probability for a delay reaction to reach completion when the system is at the fixed point. For the gestation model this is the fraction of pregnant individuals which successfully give birth, for the juvenile model this is the fraction of juveniles which successfully mature and reach the reproductive age.

\begin{figure}[t!!]
  \centering
    \includegraphics[width=0.3\textwidth,angle = 270]{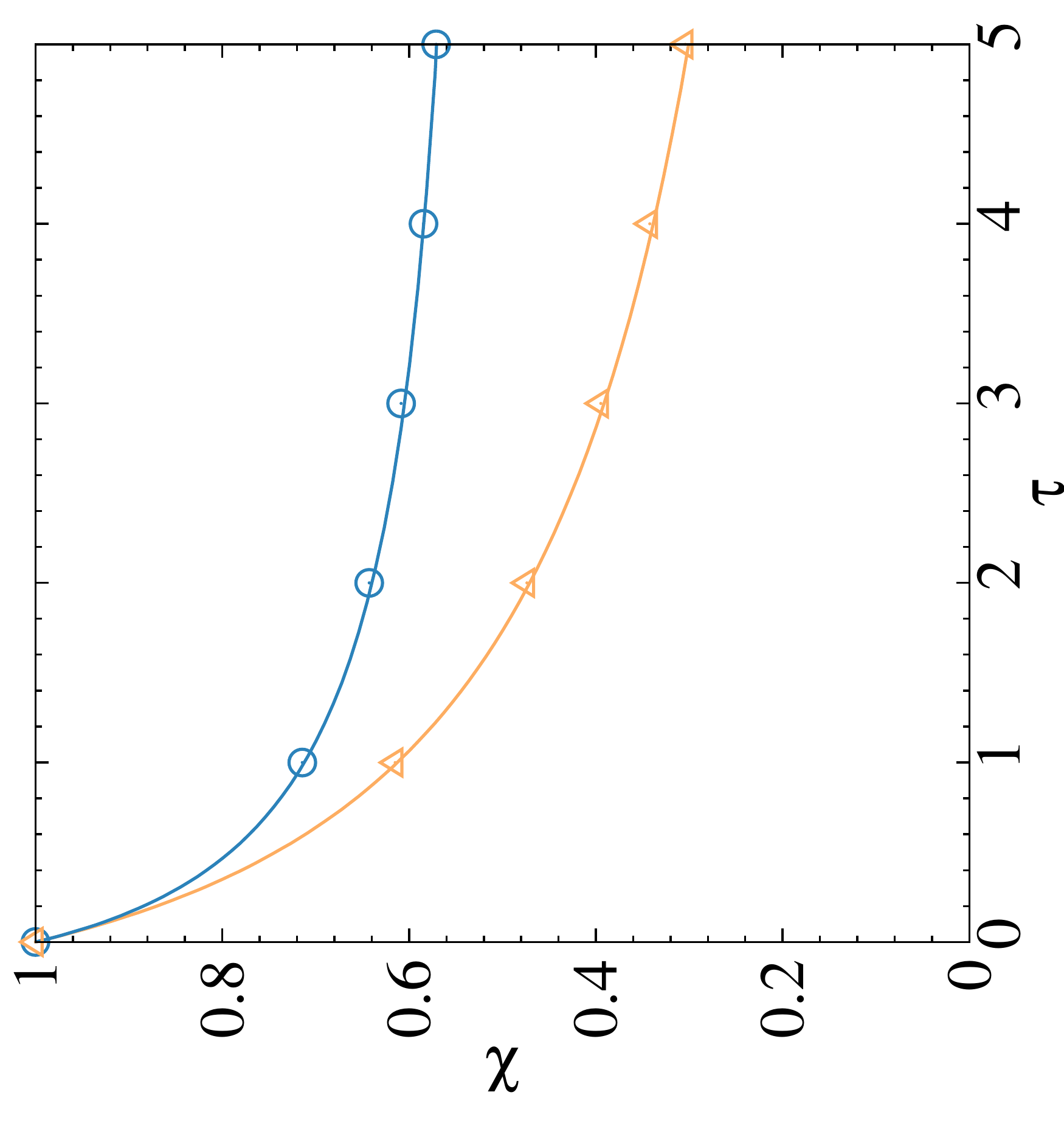}
 \caption{Relationship between survival probability at the fixed point, $\chi$, and delay period, $\bar{\tau}$. Lines are parametric plots using Eq.~\eqref{eq:pp_delay}. Blue lines and circles correspond to the gestation model, yellow lines and triangles to the juvenile model. Parameters are: $b = p = k = 1$ and $d=0.2$. For the gestation model the survival probability approaches $0.5$, whereas for the juvenile model it approaches 0. Symbols are the fraction of delay reactions which completed in simulations using the MNRM, averaged over 100 realisations and with $\Omega = 1000$.}
\label{fig:fig:prey_survival_prob}
\end{figure}

There are three solutions to the above fixed point equations: all extinct ($x=x_{\rm tot} = y = x' = 0$), only prey ($x=x_{\rm tot} = k$, $y = x' = 0$), and coexistence. We will focus on the coexistence fixed point, and we will simply refer to it as the fixed point. We see that $x_{\rm tot}^* = d/p$ for both models and for any delay distribution. The coexistence fixed point only exists if $d/p < k$. If $d/p \ge k$ then $h(d/p) = (1- \frac{d}{pk}) \le 0$ and Eqs.~\eqref{eq:pp_eom_x_fp}--\eqref{eq:pp_eom_xp_fp} have no consistent solutions. If $d/p<k$ we can rearrange Eqs.~\eqref{eq:pp_eom_x_fp}--\eqref{eq:pp_eom_xp_fp}, and find
\begin{align}
y^* &= p^{-1}[w_x\chi(y^*)  + v_x ]h(\tfrac{d}{p}),\label{eq:pp_eom_x_fp_rearranged}  \\
x^* &= \phi^{-1}[w_x \chi(y^*)  + v_x ]\frac{d}{p}, \label{eq:pp_eom_y_fp_rearranged} \\
x'^* &= \phi^{-1}[1-\chi(y^*)]\frac{d}{p}, \label{eq:pp_eom_xp_fp_rearranged}
\end{align}
with $\phi = w_x\chi(y^*)  + v_x + 1 - \chi(y^*)$. We will now restrict the further discussion to the case of constant delay $\bar\tau$, i.e.  $K(\tau) = \delta(\tau - \bar{\tau})$. In this case we can proceed with the analysis, and find $\chi(y^*) = e^{-py^*\bar{\tau}}$. Eqs.~\eqref{eq:pp_eom_x_fp_rearranged}--\eqref{eq:pp_eom_xp_fp_rearranged} are a transcendental set of equations for $x^*, x'^*$ and $y^*$, which cannot easily be simplified any further. However we can reparametrise the model, and treat $\chi$ as a parameter in Eqs.~\eqref{eq:pp_eom_x_fp_rearranged}--\eqref{eq:pp_eom_xp_fp_rearranged}. The delay period, $\bar\tau$, is then a function of the model parameters through $\bar{\tau} = -\ln(\chi)/(py^*)$. After substituting in for $y^*$ we find
 \begin{equation}
\bar{\tau} = -\ln(\chi)[(w_x \chi  + v_x ) h(\tfrac{d}{p})]^{-1}.
\label{eq:pp_delay}
\end{equation}
If $\chi = 1$ then there is no delay ($\bar{\tau}=0$) in either model. Furthermore $\chi$ decreases with increasing delay $\bar{\tau}$.  In the gestation model $\chi \to 0.5$ as $\bar{\tau} \to \infty$, whereas  $\chi \to 0$ in the juvenile model, as shown in Fig.~\ref{fig:fig:prey_survival_prob}. We stress that we have not formally established stability of fixed point. However for the parameters used in Fig.~\ref{fig:fig:prey_survival_prob} the fixed point has numerically been seen to be stable up to at least $\bar{\tau} = 5$.

In Fig. \ref{fig:fig:prey_survival_prob} we compare these theoretical predictions against data from simulations of the microscopic dynamics in finite populations ($\Omega=1000$). As seen in the figure the above deterministic analysis is in good agreement with stochastic simulations of the microscopic process in a finite system, $\Omega = 1000$.

\begin{figure}[t!!]
  \centering
    \includegraphics[width=0.3\textwidth,angle = 0]{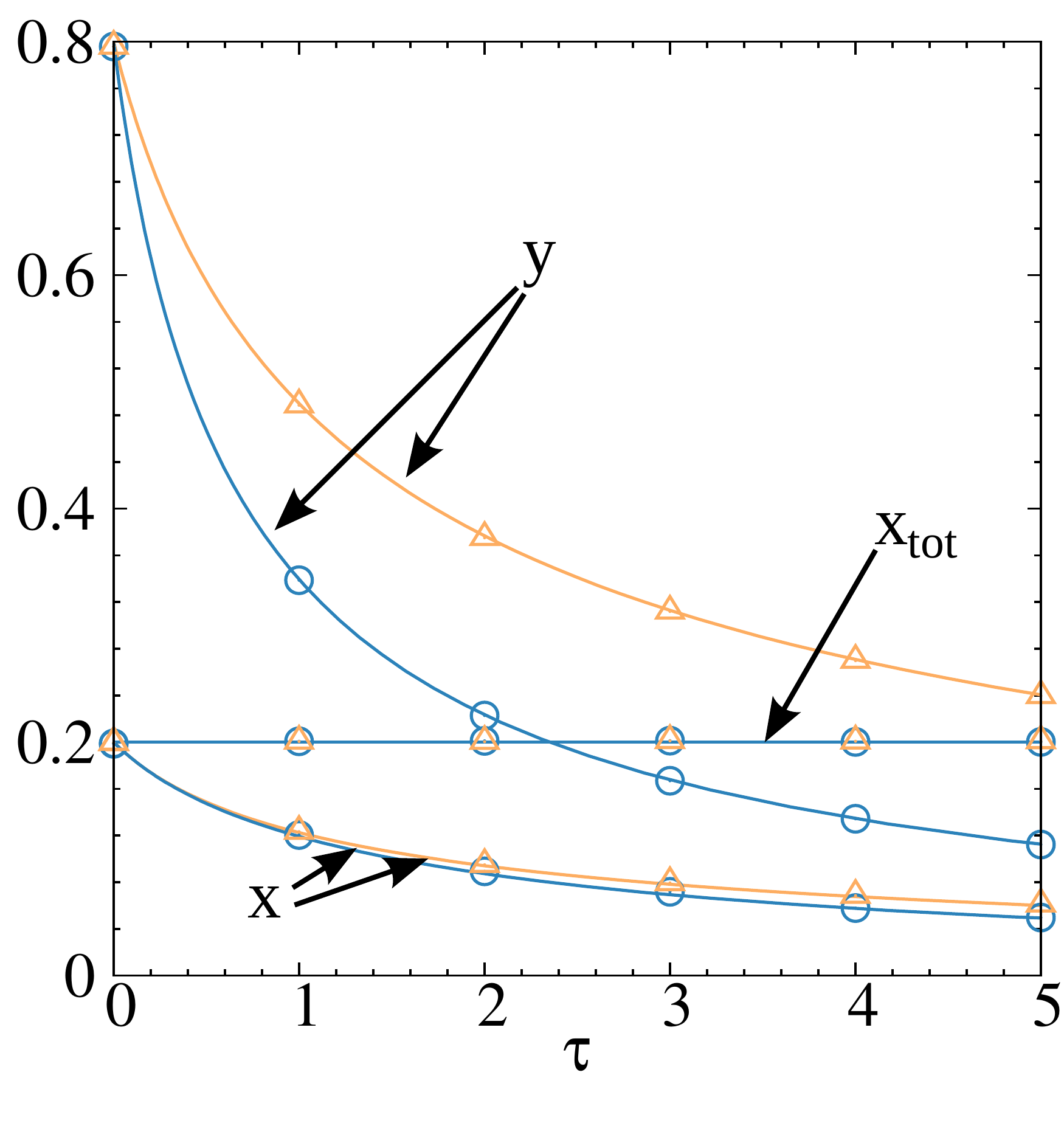}
 \caption{Relationship between fixed point location and delay period, $\bar{\tau}$. Lines are parametric plots using Eq.~\eqref{eq:pp_delay} and Eqs.~\eqref{eq:pp_eom_x_fp_rearranged}--\eqref{eq:pp_eom_xp_fp_rearranged}. Blue lines and circles correspond to the gestation model, yellow lines and triangles to the juvenile model. Parameters are: $b = p = k = 1$ and $d=0.2$. The total prey concentration, $x_{\rm tot}^*$  is not affectected by the delay (black line), and is the same for both models. Symbols are from simulations using the MNRM averaged over 100 realisations with $\Omega = 1000$. }
\label{fig:fp_with_tau}
\end{figure}

The concentration of predator and prey individuals $x^*$ and $y^*$, determined by Eqs.~\eqref{eq:pp_eom_x_fp_rearranged}--\eqref{eq:pp_eom_xp_fp_rearranged} decrease with the delay period in both models, whereas the concentration of intermediary individuals, $x'^*$, naturally increases when the delay period becomes longer. This is summarised in Fig.~\ref{fig:fp_with_tau}, obtained as a parametric plot of Eqs.~\eqref{eq:pp_eom_x_fp_rearranged}--\eqref{eq:pp_eom_xp_fp_rearranged} and Eq.~\eqref{eq:pp_delay}. Comparison with exact stochastic simulations at finite $\Omega$ again shows good agreement. We notice that the concentration of predators is much more sensitive to the choice of model than the concentration of prey for non-zero delay.

\subsection{Linear-noise approximation}

\begin{figure}[t]
  \centering
    \includegraphics[width=0.4\textwidth,angle = 270]{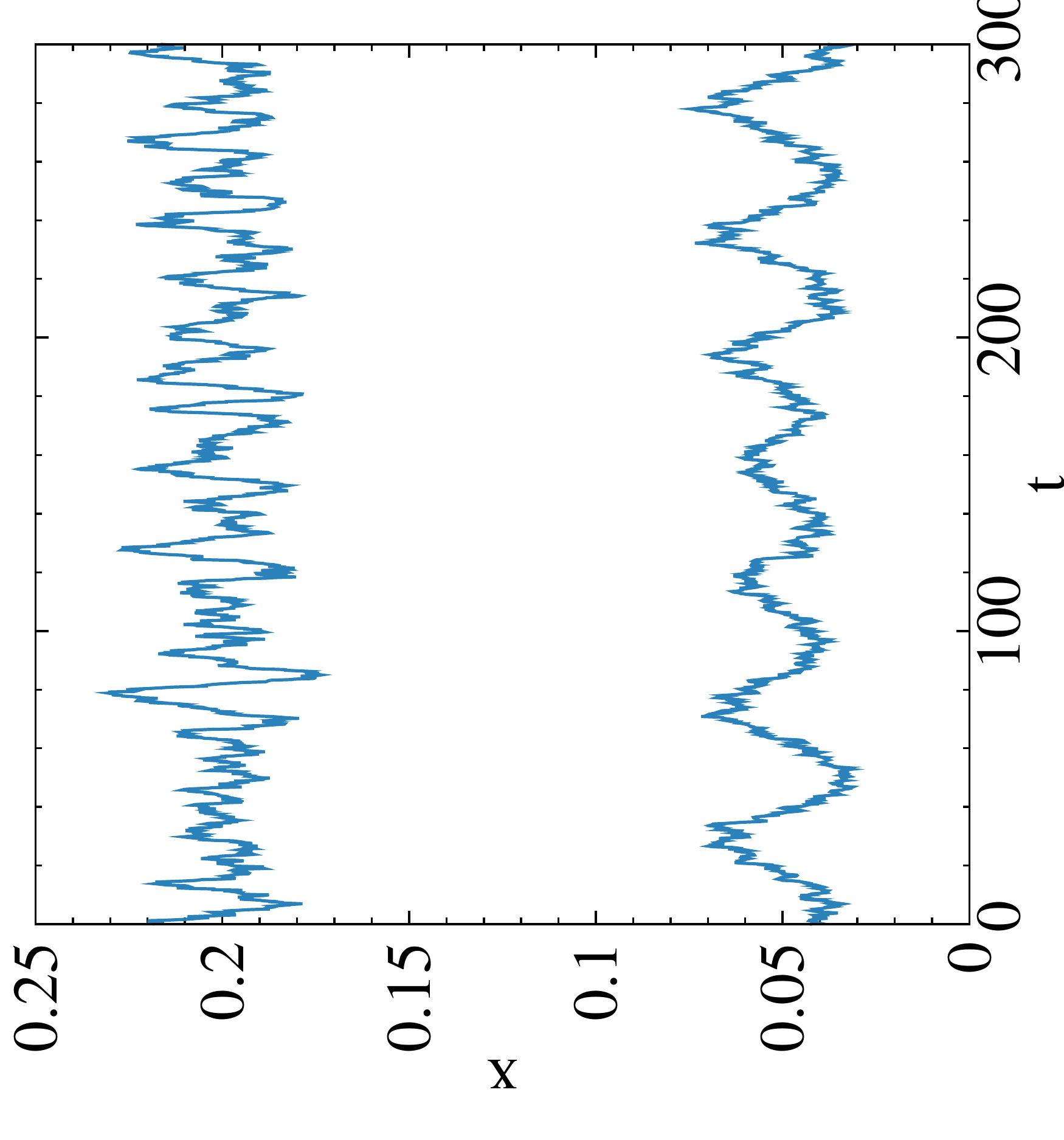}
 \caption{Sample trajectories of $x$ using the gestation model for $\bar{\tau} = 0$ (top line) and $\bar{\tau} = 5$ (bottom line). Parameters are: $b = p = k = 1$ and $d=0.2$. Simulations performed using the MNRM with $\Omega = 10000$. Both trajectories are observed to oscillate about their respective deterministic fixed points. The  oscillations for $\bar{\tau}=5$ are much more defined, with a period $T \sim 50$.}
\label{fig:sample_traj}
\end{figure}

The above deterministic analysis is valid only in the limit of infinite populations. We next study the effects of noise induced by the stochastic dynamics when the number of individuals in the system is finite. Representative trajectories generated from stochastic simulations in Fig.~\ref{fig:sample_traj} show that the effects of noise can be quite profound. As seen in numerous Markovian systems \cite{mckane,alonso.2007,biancalani.2010,biancalani.2011,andyalan} noise can generate sustained oscillations in parameter regimes in which the purely deterministic system approaches a stable fixed point. This effect has also been observed in stochastic dynamics with fixed and distributed delay \cite{barrio,galla,brett_galla,brett_galla2014}, the time series shown in Fig. \ref{fig:sample_traj} show that this phenomenon extends to delay dynamics with uncertain completion of delay reactions.

\begin{figure}[t]
  \includegraphics[width=0.4\textwidth,angle = 0]{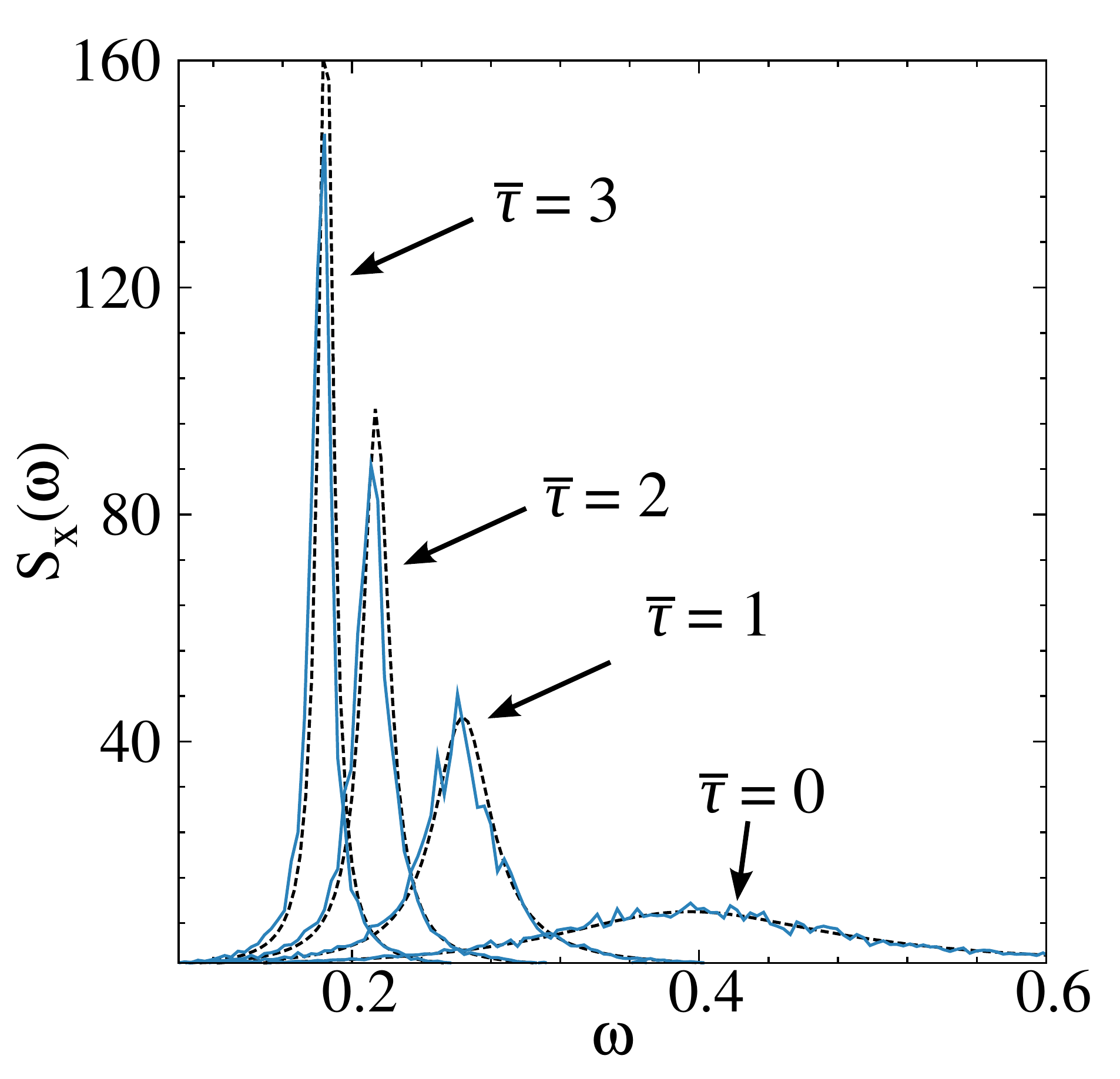} \includegraphics[width=0.4\textwidth,angle = 0]{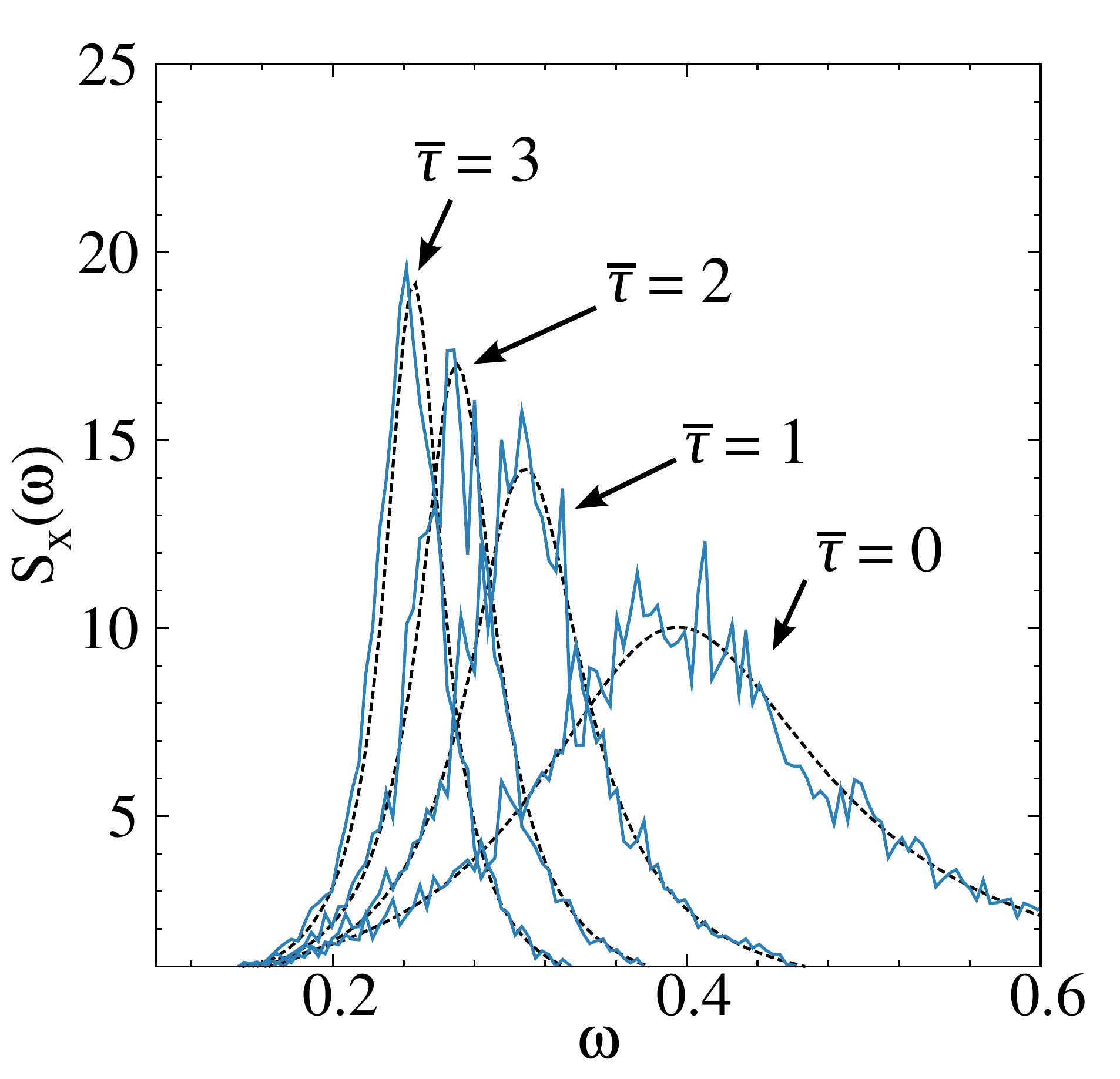}
 \caption{Variation of the power spectrum of $x$, $S_{x}(\omega)$ with the delay period, $\bar{\tau}$. Top panel corresponds to the gestation model, bottom panel to the juvenile model. Dashed lines are theoretical predictions using Eq.~\eqref{eq:pp_delay}, solid lines are from simulations using the MNRM with $\Omega = 10000$. Parameters are: $b = p = k = 1$ and $d=0.2$.  }
\label{fig:ps}
\end{figure}

The oscillations can be characterised by the power spectra of fluctuations about the deterministic fixed point, more precisely by the diagonal elements od $\uu{S}(\omega) = \avg{\widetilde{\bm{\xi}}(\omega)\widetilde{\bm{\xi}}^\dagger(\omega)}$, introduced before Eq. (\ref{eq:spectrum_matrix}). Fig.~\ref{fig:ps} shows how the power spectrum responds to an increase in the delay period. In both models the effect of delay is to increase the period of the quasi-cycles, as can be seen by a shift of the peak towards lower frequencies with increasing delay. Delay also increases the amplitude of the oscillations. For the gestation model the effect is more pronounced, the amplitude is much more sensitive to the delay.

To calculate the power spectrum of fluctuations analytically we start from the general expression for the Fourier transform of the Langevin equation found in the linear-noise approximation, Eq.~\eqref{eq:interrupt_lna_langevin_ft}. For the predator-prey models we have
\begin{align}
i\omega \widetilde\xi_x(\omega) =&~ \Big(h(x_{\rm tot}^*) -\frac{bx^*}{k} - py^*\Big)\Big(v_x+\widetilde L_x(\omega) \Big) \widetilde\xi_x(\omega) \nonumber \\ 
&-\frac{bx^*}{k}\Big(v_x+\widetilde L_x(\omega)\Big) \widetilde \xi_{x'}(\omega) \nonumber \\
&- px^*\Big(1+h(x_{\rm tot}^*)\Big)G_x(\omega)\widetilde\xi_y(\omega) + \widetilde\eta_x(\omega) , \label{eq:pp_ft_lna_x} \\
i\omega\widetilde\xi_{x'}(\omega)  =&~\Big(h(x_{\rm tot}^*)-\frac{bx^*}{k}\Big)\Big(1+\widetilde L_{x'}(\omega)\Big)\widetilde\xi_x(\omega) \nonumber \\ &-\frac{bx^*}{k}\Big(1+\widetilde L_{x'}(\omega)\Big)\xi_{x'}(t) \nonumber \\
&- \Big(px'^*+ px^*h(x_{\rm tot}^*)\Big)G_{x'}(\omega)\xi_y(t) + \widetilde\eta_{x'}(\omega),  \label{eq:pp_ft_lna_xprime} \\
i\omega\widetilde \xi_y(\omega) =&~ \Big(py^* -\frac{bx^*}{k}\Big)\widetilde L_y(\omega) \widetilde\xi_x(\omega)  -\frac{bx^*}{k}\widetilde L_y(\omega)\widetilde \xi_{x'}(\omega) \nonumber \\
&- px^*h(x_{\rm tot}^*) G_y(\omega) \xi_y(t)  + \widetilde\eta_y(\omega) ,
\label{eq:pp_ft_lna_y}
\end{align}
where $\widetilde L_\alpha(\omega) = \int_0^\infty d\tau~ e^{-i\omega \tau
} L_\alpha(\tau)$ and $G_\alpha(\omega) = \frac{1}{i\omega}\int_0^\infty d\tau~ L_\alpha(\tau)\left\{1-e^{-i\omega\tau}\right\}$. For these models the weighted delay effects are $\bm{L}(\tau) = [w_x  K(\tau), -(K(\tau) +  Q(\tau)), Q(\tau)]$.

Eqs.~\eqref{eq:pp_ft_lna_x}--\eqref{eq:pp_ft_lna_y} can be rewritten as a matrix equation $\widetilde\boldeta(\omega) = \uu{M}(\omega)\widetilde \boldxi(\omega)$. The noise correlation matrix has elements
\BE
\widetilde B_{x,x}(\omega) &=& c \Big(v_x^2+\chi w_x^2 + w_x \chi  \nonumber \\
&&+ v_x + 2v_xw_x\text{Re}\widetilde K(\omega)\Big), \nonumber \\
\widetilde B_{x,x'}(\omega) &= &c \Big(v_x -\chi w_x+w_x\widetilde K(\omega) \nonumber \\
&&-v_x(\widetilde K^*(\omega)+\widetilde Q^*(\omega)) \Big), \nonumber \\
\widetilde B_{x,y}(\omega) &=& c \left(v_x\widetilde Q^*(\omega) -w_x \chi  - v_x    \right) ,\nonumber \\
\widetilde B_{x',x'}(\omega) &=& 2c \left(1 - \text{Re}(\widetilde K(\omega)+\widetilde{Q}(\omega) )\right), \nonumber \\
\widetilde B_{x',y}(\omega) &=& c \left(  \widetilde{Q}^*(\omega)-(1-\chi) \right), \nonumber \\
\widetilde B_{y,y}(\omega) &=& 2c\phi ,
\label{eq:pp_B_matrix}
\EE
where $c = x^*h(x_{\rm tot}^*)$. The remaining elements follow from the Hermitian property,  $\uu{\widetilde B}(\omega) = \uu{\widetilde B}^\dagger(\omega)$. The spectrum matrix $\uu{S}(\omega)$ can be found by inserting Eqs.~\eqref{eq:pp_ft_lna_x}--\eqref{eq:pp_B_matrix} into Eq.~\eqref{eq:spectrum_matrix}. Comparison of these theoretical predictions against data from simulation shows good agreement, see Fig. \ref{fig:ps}, and confirms the viablity of the linear-noise approximation for both models.


If we return back to the definition of the models, the delay reaction for the gestation model is
\begin{align}
 X &\overset{b(1-x_{\rm tot}/k)}\longrightarrow X' \,;\, X' \overset{K(\tau)}\Longrightarrow 2X ,\label{eq:d1}
\end{align}
whereas for the juvenile model it is
\begin{align}
 X &\overset{b(1-x_{\rm tot}/k)}\longrightarrow  X+ X' \,;\, X' \overset{K(\tau)}\Longrightarrow X.\label{eq:d2}
\end{align}
The total effect of the delay reaction (i.e. of the initial effects together with the delayed effects) is the same for both models: the number of individuals of type $X$ increases by one. The difference between the models is in the intermediary changes.

To probe the effects of delay, let us compare the outcome of the delay models ($\bar\tau>0$) with the corresponding Markovian models (obtained in the limit $\bar{\tau} \to 0$). In this limit the reactions in Eq. (\ref{eq:d1}) and Eq. (\ref{eq:d2}) both reduce to
\begin{align}
  X &\overset{b(1-x/k)}\longrightarrow 2X,
\end{align}
The study of the fixed point behaviour and of the power spectrum of the models with delay shows that the location of the fixed point and the peak of the power spectrum at any fixed $\bar\tau>0$ are closer to the corresponding quantities at $\bar\tau=0$ in the juvenile model than in the gestation model. This indicates that it is not only the duration of the delay which determines whether or not it can be neglected, but also the details of the delay reaction. We note that the survival probability at the fixed point (shown in Fig.~\ref{fig:fig:prey_survival_prob}) is more sensitive to the delay in the juvenile model than in the gestation model.
\section{Conclusions}

The main focus of this paper has been to extend the generating-functional approach for stochastic interacting particle systems to dynamics in which delay reactions can be interrupted. In particular we consider cases in which the interruption rate depends on the state of the system at the time of interruption. The probability with which the delay effects ultimately occur is then a functional of the path taken by the system during the delay period. We successfully set up a path-integral description of such systems.

The action of the generating functional for delay reactions with interruption has a clear relationship with the actions found previously in models in which delay reaction complete with certainty. The example in Sec.~\ref{sec:interrupt_sir} shows how our extended theory includes results previously derived in simpler cases where the interruption rate is constant and independent of the state of the system at the time of interruption. The generating functional provides a starting point for further analytical calculations, specifically we derive Gaussian and linear-noise approximations from it to characterise noise-induced effects in finite populations.

We apply these techniques to an example inspired by predator-prey dynamics. We study two variants of the model, one with gestation delay and one with maturation delay. The deterministic approximation can be used to give a good estimate of the probability that any given delay reaction completes. In the gestation model this describes the probability that a pregnant individual successfully gives birth, in the juvenile model it is the probability that a newborn individual successfully reaches adulthood.  Starting from the linear-noise approximation we show how the formalism can be used to analytically study persistent noisy cycles, induced by demographic stochasticity. 

Finally, the overall effect of delay reactions is identical in the gestation and in the juvenile predator-prey model, that is to say the changes of delay reactions at the time they trigger taken together with the effects at completion are the same in both models. Our analysis shows that different quantities (e.g. survival probability, fixed point location, peak of the power spectrum) have different sensitivities to the delay in the two models. For both the deterministic fixed point and the characteristic frequency of noise-induced oscillations, the duration of the delay period has a much stronger effect on the outcome in the gestation model than in the juvenile model. Conversely, the probability that the delayed effects successfully occur decreases much faster with increasing delay in the juvenile model than in the gestation model.

These observations have implications for the construction of population models. Whether a delay can safely be neglected depends not just on the duration of the delay period itself but also on the details of the changes to the population at the start and end of the delay period. This indicates that care must be taken when approximating models with delay or intermediate processes by effective Markovian dynamics.

In summary, our previous work \cite{brett_galla} together with the present paper provides a systematic and coherent approach to studying stochastic effects in a wide range of delay systems, including those with distributed delay, delay reactions that can fail to complete with path-dependent rates and systems with multiple possible interruption channels. The generating functional is the natural mathematical entity which with to address these systems, and is the analog of the master equation in the context of Markovian dynamics. We have here only applied this method to a small set of examples, inspired by ecological dynamics, but we anticipate that the formalism that is now in place can be of value in the context of many other applications, including those in epidemiology, metabolic dynamics, gene expression and other fields which delayed dynamics.

\begin{acknowledgements}
TB would like to thank the EPSRC (UK) for support.
\end{acknowledgements}

\onecolumngrid
\begin{appendix}

\section{The generating functional for interruptible delay}

\label{sec:gf_appendix}
\subsection{Conditional probabilities}

To derive the generating functional we focus on the model in discretised time. The continuous-time limit is taken at the end of the calculation. We adopt the following notation: $\bm{k}_t$ denotes all $k_{i,t'}^\tau$ for which $t'=t$, $\bm{\ell}_t$ denotes all $\ell_{i,t'}^{\tau,s}$ for which $t'+s=t$, and $\bm{m}_t$ denotes all $m_{i,t'}^\tau$ for which $t'+\tau=t$.

Using this notation the equation of motion for $x_{t,\alpha}$ can be expressed in the form $x_{t+\Delta,\alpha} = x_{t,\alpha} + g_\alpha(\bm{k}_t,\bm{\ell}_t,\bm{m}_t)$. The path probability $P(\bm{x},\bm{k},\bm{\ell},\bm{m})$ can be written as the product
\be
	P(\bm{x},\bm{k},\bm{\ell},\bm{m}) = \prod_t  P(\bx_{t+\Delta}| \{\bx,\bm{k},\bm{\ell},\bm{m}\}_{t'\le t}) P(\bm{k}_t, \bm{\ell}_t,\bm{m}_t|\bm{x}_t,  \{\bx,\bm{k},\bm{\ell},\bm{m}\}_{t'<t}).
\ee
For a particular $t$, all of the variables contained in $\{\bm{k}_t, \bm{\ell}_t,\bm{m}_t\}$ are independent of each other (conditioned on the history of the system up to time $t$), so their joint probability distribution factorises
\BE
	P(\bm{k}_t, \bm{\ell}_t,\bm{m}_t| \bm{x}_t,  \{\bx,\bm{k},\bm{\ell},\bm{m}\}_{t'<t}) &=& \prod_i\prod_{\tau >0}\bigg\{P(m_{i,t-\tau}^\tau|\bx_t, \{\bx,\bm{k},\bm{\ell},\bm{m}\}_{t'< t}) \prod_{0<s<\tau} P(\ell_{i,t-s}^{\tau,s}|\bx_t, \{\bx,\bm{k},\bm{\ell},\bm{m}\}_{t'< t})\nonumber \\
	&&\times P(k_{i,t}^\tau|\bx_t, \{\bx,\bm{k},\bm{\ell},\bm{m}\}_{t'< t})\bigg\}.
	\label{eq:path_cond_prob}
\EE
All of these conditional probabilities are known from the definition of the process. We have
\begin{align}
	P(m_{i,t-\tau}^\tau|\bx_t, \{\bx,\bm{k},\bm{\ell},\bm{m}\}_{t' < t}) &=  P(m_{i,t-\tau}^\tau |\{\ell_{i,t-\tau}^{\tau,\sigma}\}_{0<\sigma<\tau}, k_{i,t-\tau}^\tau) \nonumber \\
	P(\ell_{i,t-s}^{\tau,s}|\bx_t, \{\bx,\bm{k},\bm{\ell},\bm{m}\}_{t' < t}) &= P(\ell_{i,t-s}^{\tau,s}|\bx_{t},\{\ell_{i,t-s}^{\tau,\sigma}\}_{0<\sigma<s}, k_{i,t-s}^\tau) \nonumber \\
	P(k_{i,t}^\tau|\bx_t, \{\bx,\bm{k},\bm{\ell},\bm{m}\}_{t' < t}) &= P(k_{i,t}^\tau|\bx_t).
	\label{eq:react_variables_prob_app}
\end{align}
The explicit expressions for the probabilities in terms of $f_i(\bx_t)$, $r_i(\bx_t)$ and $K_i(\tau)$ are given by Eqs.~\eqref{eq:prob_k}--\eqref{eq:prob_m_1}. The probability distribution $ P(\bx_{t+\Delta}| \{\bx,\bm{k},\bm{\ell},\bm{m}\}_{t'\le t})$ is the product of $\delta$-distributions given in Eq.~\eqref{eq:delta_eom}, which for clarity we repeat here,
\begin{align}
	P(\bx_{t+\Delta}| \{\bx,\bm{k},\bm{\ell},\bm{m}\}_{t'\le t}) =\prod_\alpha \delta\bigg(x_{t+\Delta,\alpha}-x_{t,\alpha} -g_\alpha(\bm{k}_t,\bm{\ell}_t,\bm{m}_t)\bigg) .
		\label{eq:delta_eom_appendix}
\end{align}
We have
\begin{align}	
	g_\alpha(\bm{k}_t,\bm{\ell}_t,\bm{m}_t)=\frac{1}{\Omega}\sum_i \sum_{\tau >0}\bigg[ v_{i,\alpha} k_{i,t}^{\tau} + w_{i,\alpha}m_{i,t-\tau}^{\tau} + \sum_{0<s< \tau} u_{i,\alpha}\ell_{i,t-s}^{\tau,s} \bigg].
\end{align}
To make the expressions that follow more compact we can define the product
\be
	\Phi_{i,t,\tau} = P(k_{i,t}^\tau|\bx_t)P(m_{i,t}^\tau |\{\ell_{i,t}^{\tau,\sigma}\}_{0<\sigma<\tau}, k_{i,t}^\tau) \prod_{0<s<\tau}P(\ell_{i,t}^{\tau,s}|\bx_{t+s},\{\ell_{i,t}^{\tau,\sigma}\}_{0<\sigma<s}, k_{i,t}^\tau),
	\label{eq:Phi_def}
\ee
so that the path probability factorises into 
\BE
	P(\bm{x},\bm{k},\bm{\ell},\bm{m}) &=& \prod_t P(\bx_{t+\Delta}| \{\bx,\bm{k},\bm{\ell},\bm{m}\}_{t'\le t}) \prod_{\tau >0, i, t}\Phi_{i,t,\tau}.
	\label{eq:path_prob_2}
\EE
We will refer to $\Phi_{i,t,\tau}$ as the statistical weight associated with the combination of random variables $\{k_{i,t}^\tau,\{\ell_{i,t}^{\tau,\sigma}\}_{0<\sigma<\tau}, m_{i,t}^\tau\}$. It is not itself a probability, as the probabilities which compose it are conditioned on different random variables.  

All of the conditional probabilities which make up $\Phi_{i,t,\tau}$ in Eq.~(\ref{eq:Phi_def}) are only conditioned on variables which are either i) $\{\bm{x}_{t'}\}_{t \le t' < t+\tau}$ or ii) other members of $\{k_{i,t}^\tau,\{\ell_{i,t}^{\tau,\sigma}\}_{0<\sigma<\tau}, m_{i,t}^\tau\}$. The weight $\Phi_{i,t,\tau}$ is therefore only a function of  $\{k_{i,t}^\tau,\{\ell_{i,t}^{\tau,\sigma}\}_{0<\sigma<\tau}, m_{i,t}^\tau\}$ and $\{\bm{x}_{t'}\}_{t \le t' < t+\tau}$. To simplify the notation we suppress these arguments.

Later on this definition of $\Phi_{i,t,\tau}$ will allow us to factorise the generating functional and perform the sums for each factor separately.

Inspecting Eqs.~\eqref{eq:prob_k}--\eqref{eq:prob_m_1} we can identify the only combinations of values of $k_{i,t}^\tau$, $\ell_{i,t}^{\tau, 0< s< \tau}$, and $m_{i,t}^\tau$ for which $\Phi_{i,t,\tau}$ is non-zero. They are as follows:\\

(i) A reaction of type $i$ with delay period $\tau$ does not fire at time $t$. In this case we have
\BE
k_{i,t}^\tau&=&0, \nonumber\\
\ell_{i,t}^{\tau,\sigma}&=&0 ~\forall ~0<\sigma<\tau,\nonumber \\
 m_{i,t}^\tau &=&0,
\EE
with
\begin{align}
\Phi_{i,t,\tau} = 1- \Delta^2K_i(\tau)\Omega r_i(\bm{x}_t).
\label{eq:prob_no_fire}
\end{align}
\\

(ii) A reaction of type $i$ with delay period $\tau$ fires at time $t$ and is interrupted at time $t+s$. The values of the random variables  for this case are
\BE
k_{i,t}^\tau&=&1,\nonumber \\
\ell_{i,t}^{\tau,\sigma}&=&0~\forall ~0<\sigma<s \nonumber \\
\ell_{i,t}^{\tau,s}&=&1, \nonumber \\
\ell_{i,t}^{\tau,\sigma}&=&0~\forall ~s<\sigma<\tau, \nonumber \\
m_{i,t}^\tau &=&0,
\EE
and
\BE
\Phi_{i,t,\tau} =&~ \Delta f_i(\bm{x}_{t+s})\prod_{0<\sigma<s}\Big(1-\Delta f_i(\bm{x}_{t+\sigma})\Big)  \Delta^2K_i(\tau)\Omega r_i(\bm{x}_t).
		\label{eq:prob_fire_inter}
\EE
\\

(iii) A reaction of type $i$ with delay period $\tau$ fires at time $t$ and is not interrupted. The random variables take the values
\BE
k_{i,t}^\tau&=&1,\nonumber \\
\ell_{i,t}^{\tau,\sigma}&=&0~\forall ~0<\sigma<\tau,\nonumber \\
m_{i,t}^\tau &=&1,
\EE
with
\begin{align}
\Phi_{i,t,\tau} = \prod_{0<\sigma<\tau}\Big(1-\Delta f_i(\bm{x}_{t+\sigma})\Big)\Delta^2K_i(\tau)\Omega r_i(\bm{x}_t).
		\label{eq:prob_fire_nointer}
\end{align}

\subsection{Generating functional}

Inserting Eq.~\eqref{eq:path_prob_2} and Eq.~\eqref{eq:delta_eom_appendix} into the definition of the generating functional gives
\begin{align}
	Z[\bm{\psi}] = \int D\bx &\sum_{\{\bm{k},\bm{\ell},\bm{m}\}} \prod_{t,\alpha} \delta\bigg(x_{t+\Delta,\alpha}-x_{t,\alpha} -\frac{1}{\Omega}\sum_i \sum_{\tau >0}\bigg[ v_{i,\alpha} k_{i,t}^{\tau} + w_{i,\alpha}m_{i,t-\tau}^{\tau} + \sum_{0<s< \tau} u_{i,\alpha}\ell_{i,t-s}^{\tau,s} \bigg]\bigg) \nonumber \\
	&\times e^{-\sum_{t,\alpha}\Delta \psi_{t,\alpha}x_{t,\alpha}} \times\prod_{\tau >0,i,t} \Phi_{i,t,\tau},
\end{align}
where we have introduced the short-hand $\int D\bx \equiv \int \prod_{t}d\bm{x}_t$. Using the exponential representation of the $\delta$-function we can write
\begin{align}
	Z[\bm{\psi}] = \int \D\bx D\hat{\bx}& \sum_{\{\bm{k},\bm{\ell},\bm{m}\}} \exp\bigg(i\sum_{t,\alpha}  \hat{x}_{t,\alpha}\bigg\{x_{t+\Delta,\alpha}-x_{t,\alpha} \nonumber \\
	&-\frac{1}{\Omega}\sum_i \sum_{\tau >0}\bigg[ v_{i,\alpha} k_{i,t}^{\tau} + w_{i,\alpha}m_{i,t-\tau}^{\tau} + \sum_{0<s< \tau} u_{i,\alpha}\ell_{i,t-s}^{\tau,s} \bigg]\bigg\}-\sum_{t,\alpha}\Delta \psi_{t,\alpha}x_{t,\alpha} \bigg)\prod_{\tau >0,i,t} \Phi_{i,t,\tau},
	\label{eq:gf_exp_rep}
\end{align}
with $\int D\hat{\bx} \equiv \int \prod_{t}\frac{d \hat{\bm{x}}_t}{2\pi}$. 
In what follows we will make the transformation $i\hat{\bx} \to \bp$. Eq.~\eqref{eq:gf_exp_rep} can be factorised to give
\BE
	Z[\bm{\psi}] &= &\int \D\bx D\bp~  e^{\sum_{t,\alpha} \left( p_{t,\alpha}(x_{t+\Delta,\alpha}-x_{t,\alpha})-\Delta \psi_{t,\alpha}x_{t,\alpha}\right)} \times  \sum_{\{\bm{k},\bm{\ell},\bm{m}\}} \prod_{\tau >0,i,t} e^{ - \frac{1}{\Omega}\sum_\alpha\Big[ v_{i,\alpha}p_{t,\alpha} k_{i,t}^{\tau} + w_{i,\alpha}p_{t,\alpha}m_{i,t}^{\tau}  \Big] } \nonumber \\
	&&\times e^{ - \frac{1}{\Omega}\sum_\alpha\Big[\sum_{0 <s<\tau} u_{i,\alpha}p_{t+s,\alpha}\ell_{i,t}^{\tau,s}  \Big]}  \Phi_{i,t,\tau}.
\EE

Before we take the average over the random variables $\bm{k}$, $\bm{l}$, and $\bm{m}$ it is useful to define
\BE
	\Psi_{i,t,\tau} &=& \sum_{k_{i,t}^\tau}\sum_{m_{i,t}^\tau}\sum_{\{\ell_{i,t}^{\tau,s}\}_{0<s<\tau}} \!\!\!\!\! e^{ - \frac{1}{\Omega}\sum_\alpha\Big[ v_{i,\alpha}p_{t,\alpha} k_{i,t}^{\tau} + w_{i,\alpha}p_{t,\alpha}m_{i,t}^{\tau} \Big]} e^{ - \frac{1}{\Omega}\sum_\alpha\Big[\sum_{0 <s<\tau} u_{i,\alpha}p_{t+s,\alpha}\ell_{i,t}^{\tau,s}  \Big]}  \Phi_{i,t,\tau} .
\EE
The quantity $\Psi_{i,t,\tau}$ is a function of $\{\bm{x}_{t'}\}_{t \le <t' < t+\tau}$, for clarity we have suppressed these arguments. In defining $\Psi_{i,t,\tau}$ we have collected all occurrences of the variables indicated in the summation, $\{k_{i,t}^\tau,\{\ell_{i,t}^{\tau,\sigma}\}_{0<\sigma<\tau}, m_{i,t}^\tau\}$. As explained beneath Eq.~(\ref{eq:path_prob_2}),  $\Phi_{i,t,\tau}$ is only a function of  $\{k_{i,t}^\tau,\{\ell_{i,t}^{\tau,\sigma}\}_{0<\sigma<\tau}, m_{i,t}^\tau\}$ and of $\{\bm{x}_{t'}\}_{t \le <t' < t+\tau}$. This construction allows us to evaluate each $\Psi_{i,t,\tau}$ separately to all other terms in the generating functional, which can be written as
\BE
Z[\bm{\psi}] &= &\int \D\bx D\bp~  e^{\sum_{t,\alpha} \left( p_{t,\alpha}(x_{t+\Delta,\alpha}-x_{t,\alpha})-\Delta \psi_{t,\alpha}x_{t,\alpha}\right)} \prod_{\tau >0,i,t}  \Psi_{i,t,\tau}.
\EE

Using Eqs.~\eqref{eq:prob_no_fire},~\eqref{eq:prob_fire_inter}, and~\eqref{eq:prob_fire_nointer} we obtain
\begin{align}
	\Psi_{i,t,\tau} = 1 + \bigg\{&\sum_{0<s<\tau}e^{ - \frac{1}{\Omega}\sum_\alpha\left[ v_{i,\alpha}p_{t,\alpha} + u_{i,\alpha}p_{t+s,\alpha} \right]} \Delta f_i(\bx_{t+\sigma})\prod_{0<\sigma<s}(1-\Delta f_i(\bx_{t+\sigma}))  \nonumber \\
	&+e^{ - \frac{1}{\Omega}\sum_\alpha\left[ v_{i,\alpha}p_{t,\alpha}  + w_{i,\alpha}p_{t,\alpha} \right]}  \prod_{0<\sigma<\tau}(1-\Delta f_i(\bx_{t+\sigma}))       -1 \bigg\} \Delta^2K_i(\tau)\Omega r_i(\bx_{t}),
\end{align}

which for small $\Delta$ can also be written as an exponential, cf. $1+\Delta f = e^{\Delta f} +\mathcal{O}(\Delta^2)$. 

Repeating this procedure for all $\Psi_{i,t,\tau}$ leaves a generating functional for the discrete-time dynamics. We can then take the continuous-time limit, and arrive at 
\begin{equation}
	Z[\bm{\psi}] =  \int D\bx D\bx ~e^{ - S[\bx,\bp]- \int \! dt\, \bm{\psi}(t) \cdot \, \bm{x}(t)},
	\label{eq:interrupt_exact_gf_app}
\end{equation}
with the action

	\begin{align}
	S[\bx,\bp] = \int dt \bigg[ \bm{p}(t)\cdot \dot{\bm{x}}(t)+&\sum_i\bigg\{\int_0^\infty d\tau \bigg(e^{ - \frac{1}{\Omega}\left[ \bm{v}_{i}\cdot \bm{p}(t-\tau) + \bm{w}_{i}\cdot \bm{p}(t) \right]} - 1 \bigg) e^{-\int_0^\tau d\sigma f_i(\bx(t-\sigma))}K_i(\tau)\Omega r_i(\bx(t-\tau)) \nonumber \\
	&\hspace{-8em}+\int_0^\infty ds\bigg( e^{ - \frac{1}{\Omega}\left[ \bm{v}_{i}\cdot \, \bm{p}(t-s)  + \bm{u}_{i}\cdot \, \bm{p}(t) \right]}-1\bigg) f_i(\bx(t))e^{-\int_0^s d\sigma f_i(\bx(t-\sigma))} \int_s^\infty d\tau K_i(\tau)\Omega r_i(\bx(t-s)) \bigg\}\bigg].
	\label{eq:interrupt_hamiltonian_app}
\end{align}

\section{Multiple interruption reactions}
\label{sec:multiple_interruption}

Although the notation is involves a number of indices, it is straightforward to see how the generating functional can be extended to the case in which any delay reaction can be interrupted in multiple different ways. If reaction $i$ is a delay reaction with which can be interrupted in $\Lambda_i$ possible ways, indexed by $\mu=1,\dots, \Lambda_i$, then by extension of Eq.~\eqref{eq:discretetime_eom},
\begin{align}
	x_{t+\Delta,\alpha}-x_{t,\alpha} =& \frac{1}{\Omega}\sum_i \sum_{\tau >0}\bigg[ v_{i,\alpha} k_{i,t}^{\tau} + w_{i,\alpha}m_{i,t-\tau}^{\tau} + \sum_{0<s< \tau,\mu=1,\dots, \Lambda_i} u_{i,\mu,\alpha}\ell_{i,\mu,t-s}^{\tau,s} \bigg].
	\label{eq:discretetime_eom_multi}
\end{align}
We also have to modify the conditional probabilities of the random variables to take into account the different interruption reactions. If $k_{i,t}^\tau = 0$ then $\ell_{i,\mu,t}^{\tau,s} = 0$ for all $s$ and $\mu$, i.e.
\begin{align}
	P(\ell_{i,\mu,t}^{\tau,s} = 0|k_{i,t}^\tau=0) &= 1 \; \forall\; 0<s<\tau\;,\;\mu=1,\dots, \Lambda_i.
\end{align}
Similarly, if $k_{i,t}^\tau = 1$ and  $\ell_{i,\mu,t}^{\tau,\sigma} = 1$ then necessarily $\ell_{i,\nu,t}^{\tau,s>\sigma} = 0$ and $m_{i,t}^\tau=0$, i.e.
\begin{align}
	P(\ell_{i,\nu,t}^{\tau,s} = 0|\ell_{i,\mu,t}^{\tau,\sigma}=1, k_{i,t}^\tau=1) &= 1 \; \forall\; \sigma<s<\tau \;,\;\nu \in \Lambda_i , \nonumber \\
	P(m_{i,t}^\tau = 0 |\ell_{i,\mu,t}^{\tau,\sigma}=1, k_{i,t}^\tau=1) &= 1.
\end{align}
If the system is in state $\bm{x}_t$ then interruption through channel $\mu$ happens with rate $\Delta f_{i,\mu}(\bm{x}_t)$, 
\begin{align}
	P(\ell_{i,\mu,t}^{\tau,s} = 1|\bm{x}_{t+s},\{\ell_{i,\nu,t}^{\tau,\sigma}=0\}_{0<\sigma<s,\; \nu=1,\dots,\Lambda_i}, k_{i,t}^\tau=1) &=\Delta f_{i,\mu}(\bm{x}_{t+s}), \nonumber \\ P(\ell_{i,\mu,t}^{\tau,s} = 0|\bm{x}_{t+s},\{\ell_{i,\nu,t}^{\tau,\sigma}=0\}_{0<\sigma<s,\; \nu \in \Lambda_i}, k_{i,t}^\tau=1) &= 1-\Delta f_{i,\mu}(\bm{x}_{t+s}).
\end{align}
If the reaction has not been interrupted then the reaction always completes,
\begin{align}
	P(m_{i,t}^{\tau} = 1|\{\ell_{i,\nu,t}^{\tau,\sigma}=0\}_{0<\sigma<\tau,\; \nu=1,\dots, \Lambda_i}, k_{i,t}^\tau=1) &=  1.
\end{align}
As we are working with small $\Delta$, 
\begin{align}
	P(\ell_{i,\mu,t}^{\tau,s} = 1,\;\ell_{i,\nu,t}^{\tau,s} = 1|\bm{x}_{t+s},\{\ell_{i,\lambda,t}^{\tau,\sigma}=0\}_{0<\sigma<\tau,\; \lambda=1,\dots,\Lambda_i}, k_{i,t}^\tau=1) = \delta_{\mu,\nu}.
\end{align}

The only non-zero combinations of these conditional probabilities are:
\begin{enumerate}
	\item Reaction does not fire:
	\begin{align}
		P(m_{i,t}^\tau = 0 | k_{i,t}^\tau=0)&\!\!\!\! \prod_{\substack{0<s<\tau \\ \mu=1,\dots, \Lambda_i}} \!\!\!\! P(\ell_{i,\mu,t}^{\tau,s} = 0|k_{i,t}^\tau=0) P(k_{i,t}^\tau = 0|\bm{x}_t) = 1- \Delta^2K_i(\tau)\Omega r_i(\bm{x}_t).
		\label{eq:prob_no_fire_multi}
	\end{align}
	\item Reaction fires and is interrupted at time $t+s$ through channel $\mu$:
	\begin{align}
		&P(m_{i,t}^\tau = 0 |\ell_{i,\mu,t}^{\tau,s} = 1, k_{i,t}^\tau=1)  \!\!\!\! \prod_{\substack{s<\sigma<\tau \\ \nu=1,\dots,\Lambda_i}} \!\!\!\! P(\ell_{i,\nu,t}^{\tau,s} = 0|\ell_{i,\mu,t}^{\tau,s} = 1,k_{i,t}^\tau=1)\nonumber \\
		& \times P(\ell_{i,\mu,t}^{\tau,s} = 1|\bm{x}_{t+s},\{\ell_{i,\nu,t}^{\tau,\sigma}=0\}_{0<\sigma<s,\; \nu=1,\dots, \Lambda_i}, k_{i,t}^\tau=1)\nonumber \\
		&\times \!\!\!\!\!\! \prod_{\substack{0<\sigma<s \\ \nu \in \Lambda_i}} \!\!\!\! P(\ell_{i,\nu,t}^{\tau,\sigma} = 0|\bm{x}_{t+\sigma},\{\ell_{i,\lambda,t}^{\tau,\rho}=0\}_{0<\rho<\sigma,\; \lambda=1,\dots, \Lambda_i}, k_{i,t}^\tau=1) P(k_{i,t}^\tau = 1|\bm{x}_t) \nonumber \\
		&~~~~=  \Delta f_{i,\mu}(\bm{x}_{t+s})   \!\! \prod_{\substack{0<\sigma<s \\ \nu=1,\dots, \Lambda_i}} \!\!\! \Big(1-\Delta f_{i,\nu}(\bm{x}_{t+\sigma})\Big)\Delta^2K_i(\tau)\Omega r_i(\bm{x}_t).
		\label{eq:prob_fire_inter_multi}
	\end{align}
	\item Reaction fires and is not interrupted:
	\begin{align}
		&P(m_{i,t}^{\tau} = 1|\{\ell_{i,\mu,t}^{\tau,\sigma}=0\}_{0<\sigma<\tau,\; \mu=1,\dots,\Lambda_i}, k_{i,t}^\tau=1)\nonumber \\
		&\times\prod_{\substack{0<\sigma<\tau \\ \mu=1,\dots, \Lambda_i}}P(\ell_{i,\mu,t}^{\tau,\sigma} = 0|\bm{x}_{t+\sigma},\{\ell_{i,\nu,t}^{\tau,\rho}=0\}_{0<\rho<\sigma,\; \nu=1,\dots,\Lambda_i}, k_{i,t}^\tau=1)\nonumber \\
		&\times P(k_{i,t}^\tau = 1|\bm{x}_t) \nonumber \\
		&=   \prod_{\substack{0<\sigma<\tau \\ \mu=1,\dots,\Lambda_i}}\Big(1-\Delta f_{i,\mu}(\bm{x}_{t+\sigma})\Big)\Delta^2K_i(\tau)\Omega r_i(\bm{x}_t).
		\label{eq:prob_fire_nointer_mutli}
	\end{align}
\end{enumerate}

Following the same procedure as above and keeping track of the additional index $\mu$, the generating functional is the same as when there is only one interruption reaction, only with the action
\BE
	S[\bx,\bp] &= &\int dt \bigg[ \bm{p}(t)\cdot \dot{\bm{x}}(t)\nonumber\\
	&& +\sum_i\bigg\{\int_0^\infty d\tau \bigg(e^{ - \frac{1}{\Omega}\left[ \bm{v}_{i}\cdot \bm{p}(t-\tau) + \bm{w}_{i}\cdot \bm{p}(t) \right]} - 1 \bigg) e^{-\sum_{\mu=1,\dots,\Lambda_i}\int_0^\tau d\sigma f_{i,\mu}(\bx(t-\sigma))}K_i(\tau)\Omega r_i(\bx(t-\tau)) \nonumber \\
	&&+\sum_{\mu=1,\dots, \Lambda_i}\int_0^\infty ds\bigg( e^{ - \frac{1}{\Omega}\left[ \bm{v}_{i}\cdot \, \bm{p}(t-s)  + \bm{u}_{i}\cdot \, \bm{p}(t) \right]}-1\bigg)  f_{i,\mu}(\bx(t))e^{-\sum_{\nu=1,\dots, \Lambda_i}\int_0^s d\sigma f_{i,\nu}(\bx(t-\sigma))}\nonumber \\
	&&\times \int_s^\infty d\tau K_i(\tau)\Omega r_i(\bx(t-s)) \bigg\}\bigg].
	\label{eq:interrupt_hamiltonian_app_multi}
\EE

\end{appendix}

\bibliography{./bibtexbib.bib}{}

\end{document}

%% file: preamble.tex
\newcommand{\be}{\begin{equation}}
\newcommand{\ee}{\end{equation}}
\newcommand{\bd}{\begin{displaymath}}
\newcommand{\ed}{\end{displaymath}}
\newcommand{\BE}{\begin{eqnarray}}
\newcommand{\EE}{\end{eqnarray}}

\newcommand{\br}{\ensuremath{\mathbf{r}}}

\newcommand{\bu}{\ensuremath{\mathbf{u}}}
\newcommand{\bv}{\ensuremath{\mathbf{v}}}
\newcommand{\bw}{\ensuremath{\mathbf{w}}}
\newcommand{\bx}{\bm{x}}

\newcommand{\bn}{\ensuremath{\mathbf{n}}}

\newcommand{\bp}{\ensuremath{\mathbf{p}}}

\newcommand{\D}{{\cal D}}

\newcommand{\boldxi}{{\mbox{\boldmath $\xi$}}}
\newcommand{\boldeta}{{\mbox{\boldmath $\eta$}}}

\newcommand{\avg}[1]{\left\langle{#1}\right\rangle}